\documentclass[journal]{IEEEtran}
\usepackage{CJKutf8}
\usepackage{mathrsfs}
\usepackage{amssymb}
\usepackage[mathcal]{euscript}
\usepackage{cite}
\usepackage{graphicx}
\usepackage{psfrag}
\usepackage[caption=false]{subfig}
\usepackage{url}
\usepackage{stfloats}
\usepackage{amsmath}
\usepackage{float}
\usepackage{array}
\usepackage[colorlinks,
                      linkcolor=blue,
                      anchorcolor=blue,
                      citecolor=blue]{hyperref}
\usepackage[figure]{hypcap}
\usepackage{amsthm}
\newtheorem{remark}{Remark}
\usepackage{cases}
\usepackage{booktabs}
\usepackage[ruled,linesnumbered]{algorithm2e}
\usepackage{color,xcolor}
\usepackage{setspace}
\usepackage{makecell}
\usepackage{lettrine}
\usepackage{multirow}
\usepackage{makecell}

\allowdisplaybreaks[4]

\ifCLASSINFOpdf

\else

\fi

\begin{document}
\begin{CJK}{UTF8}{gbsn}
%
\title{Secrecy Energy Efficiency Maximization in IRS-Assisted {VLC} MISO Networks with RSMA:\\ A DS-PPO approach}

\author{Yangbo Guo,~Jianhui Fan,~Ruichen Zhang,~\IEEEmembership{Member,~IEEE},~Baofang Chang,\\~Derrick Wing Kwan Ng,~\IEEEmembership{Fellow,~IEEE},~Dusit Niyato,~\IEEEmembership{Fellow,~IEEE},~and Dong In Kim,~\IEEEmembership{Fellow,~IEEE}

\thanks{Y. Guo, J. Fan, and B. Chang are with the School of Computer and Information Engineering, Henan Normal University, Xinxiang 453007, China, with the Engineering Lab of Intelligence Business and Internet of Things, Henan 453007, China (e-mail: guoyangbo@bjtu.edu.cn, 2308183005@stu.htu.edu.cn, changbaofang@htu.edu.cn).}
\thanks{R. Zhang and D. Niyato are with the College of Computing and Data Science, Nanyang Technological University, Singapore (e-mail: ruichen.zhang@ntu.edu.sg, dniyato@ntu.edu.sg).}
\thanks{D. W. K. Ng is with the School of Electrical Engineering and Telecommunications, University of New South Wales, NSW 2052, Australia (e-mail: w.k.ng@unsw.edu.au). }
\thanks{D. I. Kim is with the Department of Electrical and Computer Engineering, Sungkyunkwan University, Suwon 16419, South Korea (email: dongin@skku.edu).}
}
                     
\maketitle

\begin{abstract}
This paper investigates intelligent reflecting surface (IRS)-assisted multiple-input single-output (MISO) visible light communication (VLC) networks utilizing the rate-splitting multiple access (RSMA) scheme. {In these networks,} an eavesdropper (Eve) attempts to eavesdrop on communications intended for legitimate users (LUs). To enhance information security and energy efficiency simultaneously, we formulate a secrecy energy efficiency (SEE) maximization problem. In the formulated problem, beamforming vectors, RSMA common rates, direct current (DC) bias, and IRS alignment matrices are jointly optimized subject to constraints on total power budget, quality of service (QoS) requirements, linear operating region of light emitting diodes (LEDs), and common information rate allocation. Due to the non-convex and NP-hard nature of the formulated problem, we propose a deep reinforcement learning (DRL)-based dual-sampling proximal policy optimization (DS-PPO) approach. {The approach leverages} dual sample strategies and generalized advantage estimation (GAE). In addition, to further simplify the design, we adopt the maximum ratio transmission (MRT) and zero-forcing (ZF) as beamforming vectors in the action space. Simulation results show that the proposed DS-PPO approach outperforms traditional baseline approaches in terms of achievable SEE and significantly improves convergence speed compared to the original PPO approach. Moreover, implementing the RSMA scheme and IRS contributes to overall system performance, {achieving approximately $19.67\%$ improvement over traditional multiple access schemes and $25.74\%$ improvement over networks without IRS deployment.}
\end{abstract}
\begin{IEEEkeywords}
Secrecy energy efficiency, rate-splitting multiple access, IRS, VLC, deep reinforcement learning.
\end{IEEEkeywords}

\section{Introduction}
\label{sec:1_introduction}
\subsection{Background}
\IEEEPARstart{W}{ITH} the continuous evolution of communication technologies and the Internet-of-Things (IoT), wireless data traffic and network services are experiencing exponential growth, leading to a serious shortage of spectrum resources in traditional radio frequency (RF) communications. Consequently, it is imperative to explore available frequency spectrums and diverse access technologies \cite{10552701,8669813,Xiaohu2021Towards}. Moreover, it is estimated that by 2025, indoor wireless data traffic will constitute at least $78\%$ of the total traffic, further exacerbating pressure on already crowded indoor RF communications\cite{9743352}.

To satisfy the stringent demand for massive frequency spectrum in next generation wireless networks, light emitting diode (LED)-based visible light communication (VLC), which exploits visible light waves for data transmission, has been recognized as a promising complementary technology to traditional RF\cite{1004299,9913998}. As a dual-purpose technology for communication and illumination, VLC offers green, ubiquitous, and cost-effective access method for wireless communications in future networks. In fact, extensive studies have demonstrated that VLC is superior to RF in several key metrics, such as bandwidth and data transmission rate\cite{8675375}.

However, practical VLC channels are more sensitive to shadowing and blockages than RF channels due to the shorter wavelength of VLC, which could lead to discontinuous signal coverage (i.e., a skip-zone situation) and inferior quality of service (QoS)\cite{9968053}. Meanwhile, intelligent reflecting surface (IRS) has attracted increasing both academic and industrial sectors due to its ability to reconfigure the wireless propagation channel proactively. Specifically, an optical IRS is composed of numerous passive reflecting elements that can independently adjust the reflection of incident signals, thereby altering characteristics of the wireless channel and increasing the degrees of freedom (DoF) in wireless communications\cite{8910627}. Leveraging the highly controllable and intelligent signal reflection capabilities of IRS, the skip-zone situation in VLC can be effectively addressed, significantly improving overall system performance.

On the other hand, owing to the broadcast nature of VLC signals and the requirement of line-of-sight (LoS) links, confidential messages intended for legitimate users (LUs) are vulnerable to intercepting by potential eavesdroppers (Eves) within the network, especially in public areas\cite{10.1145/2801073.2801075}. Fortunately,  physical layer security (PLS) has attracted extensive attention from both academia and industry, and has been regarded as a promising technology to guarantee highly secure communication\cite{10.1145/2801073.2801075}. In particular, IRS also plays an important role in strengthening PLS by meticulously reshaping the wireless channel environment. Thus, with the assistance of IRS, not only the spectral efficiency but also the information security can benefit\cite{9424177,9614037,7973146}.

Despite the rich potential performance enhancement, the implementation of IRS will also introduce more transmission links, which results in more severe inter-user interference (IUI) in multi-user scenarios. In response to this critical challenge, rate-splitting multiple access (RSMA) has been widely regarded as a promising multiple access scheme. In particular, RSMA serves as a generalized framework bridging spatial division multiple access (SDMA) and non-orthogonal multiple access (NOMA)\cite{8846706}. In RSMA, each user's message is split into a common part and a private part. Specifically, the common parts are encoded jointly into a common stream, while the private parts are separately encoded into private streams. At the users' side, the common stream is first decoded and removed by successive interference cancellation (SIC), allowing each user to decode their respective private stream\cite{9831440,10301507,10032267}. 

\subsection{Related Work}
In recent years, research on IRS-assisted RSMA networks has proliferated in RF communications \cite{10109654,9912342,10233705,10510891}. For instance, in \cite{10109654}, a downlink IRS-assisted half-duplex cooperative network was investigated, where the total power consumption of the network was minimized. Also, in \cite{9912342}, an IRS-assisted uplink RSMA system was studied, where the transmit power of each user and the passive beamforming at the IRS were jointly optimized to explore the performance limits of uplink system throughput. Besides, \cite{10233705} investigated a simultaneous transmitting and reflecting (STAR)-IRS-aided networks, proposing a PPO-based approach to maximize the sum rate of the system under QoS threshold and power budget constraints. Furthermore, a fair transmission strategy for IRS-assisted RSMA systems was designed in \cite{10510891}, where the power allocation, beamforming, and decoding order were jointly optimized by alternating optimization (AO) algorithm. However, due to the non-negative and real-valued nature of visible light signals, as well as the distinct channel models between VLC and RF systems, the obtained results from the aforementioned studies cannot be directly applied to VLC systems\cite{9662064}. 

\begin{table*}[htb!]
\centering
\scriptsize
\caption{Comparison of the Proposed Work With Other Existing Works in VLC Networks}\label{tbl:related_works}
\begin{tabular}{|c|c|c|c|c|c|c|c|c|c|c|c|c|}
  \hline
  \multirow{2}{*}{Reference} & \multirow{2}{*}{IRS} & \multirow{2}{*}{Security} & \multirow{2}{*}{\makecell{Beam-\\forming}} & \multicolumn{3}{c|}{Access scheme} & \multicolumn{5}{c|}{Objective function} & \multirow{2}{*}{\makecell{DRL-based \\ approach}} \\ \cline{5-12}
                                              &                                     &                                    &                                                  & TDMA      & NOMA        & RSMA       & Sum rate & Secrecy rate & Harvested energy & EE & SEE  
  & \\ \hline
  \cite{10047999} & \checkmark &                   &                   &                   & \checkmark &                   & \checkmark &                   &  &  &  &\\ \hline
  \cite{9714890}   & \checkmark &                   &                   & \checkmark &                   &                   & \checkmark &                   &  &  &  &\\ \hline
  \cite{10299678} & \checkmark &                   &                   & \checkmark &                   &                   &                   &                   &  &  &  & \checkmark\\ \hline              
  \cite{9756553}   & \checkmark & \checkmark &                   &                   &                   &                   &                   & \checkmark &  &  &  &\\ \hline
  \cite{10236455} & \checkmark &                   &                   &                   &                   &                   &                   &                   & \checkmark &  &  &\\ \hline
  \cite{10477386} &                   &                   & \checkmark &                   &                   & \checkmark &                   &                   & \checkmark &  &  &\\ \hline
  \cite{9226406}   &                   &                   &                   &                   &                   & \checkmark & \checkmark &                   &  &  &  &\\ \hline
  \cite{9693949}   &                   &                   & \checkmark &                   &                   & \checkmark &                   &                   &  & \checkmark &  &\\ \hline
  \cite{9493176}   &                   & \checkmark & \checkmark &                   &                   &                   &                   &                   &  & \checkmark &  &\\ \hline
  \cite{9463422}   &                   & \checkmark & \checkmark &                   &                   &                   &                   &                   &  &  & \checkmark &\\ \hline
  \cite{9178765}   &                   & \checkmark & \checkmark &                   & \checkmark &                   &                   & \checkmark &  &  &  &\\ \hline
  \cite{10684100} &                   & \checkmark & \checkmark &                   &                   &                   &                   & \checkmark &  &  &  & \checkmark \\ \hline
  \makecell{Proposed \\ work}  & \checkmark & \checkmark & \checkmark &                   &                   & \checkmark &                   &                   &  &  & \checkmark & \checkmark \\ 
  \hline
\end{tabular}
\end{table*}

To date, the significant potential of deploying IRS to enhance VLC system performance has sparked extensive research. For example, \cite{10047999} investigated an IRS-assisted VLC NOMA system, where the achievable sum rate was maximized by optimizing the passive beamforming at the IRS. Also, in \cite{9714890}, the spectral efficiency of an IRS-assisted VLC system was iteratively maximized through the frozen variable algorithm and minorization-maximization algorithm, while considering practical constraints on power budget, QoS threshold, and illumination. Moreover, in \cite{10299678}, a two-stage resource management framework was proposed to maximize the user fairness under the total power consumption and QoS constraints. Besides, an IRS-assisted VLC secure system was modeled in \cite{9756553}, and the IRS configuration matrix was optimized by the iterative Kuhn-Munkres algorithm to maximize the secrecy sum rate. Furthermore, in \cite{10236455}, an IRS-assisted simultaneous lightwave information and power transfer (SLIPT) VLC system was investigated, where the amount of harvested energy was maximized while ensuring QoS requirements. However, the aforementioned studies, i.e., \cite{10047999,9714890,10299678,9756553,10236455}, only investigated the combination of IRS and VLC without incorporating RSMA.

As a matter of fact, the RSMA's ability to efficiently manage interference has spurred scholarly interest in RSMA-based VLC networks. For instance, \cite{10477386} investigated a RSMA-based SLIPT VLC system, where the precoding matrix and DC bias vector were jointly optimized to maximize the minimum harvested energy of IoT devices. Also, in \cite{9693949}, the energy efficiency of the single-cell and the multi-cell RSMA-based VLC systems was maximized exploiting a two-layer Dinkelbach's algorithm in an iterative manner, respectively. Besides, in \cite{9226406}, the superiority of RSMA to SDMA and NOMA in MISO VLC systems was discussed, where the rate splitting factor and the beamforming vectors were jointly optimized to maximize the weighted sum rate.

\subsection{Motivation and Contributions}
Although the potential of IRS and RSMA in VLC systems has been explored to some extent in the aforementioned literatures, e.g., \cite{10047999,9714890,10299678,9756553,10236455,10477386,9693949,9226406}, these two technologies were separately investigated. Indeed, the benefits achieved by integrating IRS and RSMA in a VLC network have not been systematically and thoroughly explored. Moreover, as discussed above, due to the broadcast nature of VLC channels, there exists a serious risk of potential information leakage in IRS-assisted VLC RSMA systems. As such, a crucial issue faced by the networks is how to design wireless resource allocation scheme to enhance information security. Therefore, this paper focuses on PLS in IRS-assisted VLC RSMA networks. \textit{To the best of the authors' knowledge, PLS in IRS-assisted VLC networks with RSMA have not been thoroughly investigated thus far}. 
Besides, results from the increasingly prominent issue of power consumption in wireless communication systems, energy efficiency (EE) has already been regarded as an important performance metric in designing resource allocation scheme\cite{9693949,9493176}. Thus, to enhance the information security and EE performance simultaneously, in this paper, we adopt the secrecy energy efficiency (SEE) as the performance metric to design IRS-assisted VLC RSMA networks. Table \ref{tbl:related_works} highlights the significance of our work compared with existing studies. The contributions of this paper are summarized as follows:

\begin{itemize}
  \item[$\bullet$] To enhance the SEE of the IRS-assisted VLC network with RSMA, we formulate an optimization problem to maximize the SEE by jointly optimizing the beamforming vectors, the DC bias, the RSMA common rates, and the alignment matrices of IRS elements. The problem has constraints on the QoS threshold, the total power budget, the common information rate allocation, and the linear operating region of LEDs.
  \item[$\bullet$] Since the formulated problem is a mixed-integer nonlinear programming (MINLP) problem involving binary programming and mixed variables, it represents a tremendous challenge to traditional convex optimization methods. Thus, {we adopt} deep reinforcement learning (DRL) to address the problem {given} its efficiency in tackling complex problems. Specifically, {we first transform} the optimization problem into a Markov decision process (MDP). {Then, we propose} a new dual-sampling proximal policy optimization (DS-PPO) approach. Moreover, the beamforming design in the action space is determined by exploiting the maximum ratio transmission (MRT) and zero-forcing (ZF) methods.
  \item[$\bullet$] Simulation results demonstrate that the proposed DS-PPO algorithm exhibits excellent convergence performance. Compared with existing baseline methods, DS-PPO significantly improves the SEE performance with a relatively low time cost. Moreover, the effects of varying parameters, such as the number of IRS elements, the QoS threshold, the total power budget, and the number of LUs, on system performance are explored. These findings provide valuable insights for the design of IRS-assisted VLC systems with RSMA.
\end{itemize}

The structure of the paper is organized as follows. Section \ref{sec:2_system_model} describes the system model of the considered IRS-assisted VLC network with RSMA. Section \ref{sec:3_problem_formulation} formulates the SEE maximization problem. Then, a DRL-based DS-PPO approach is presented and analysed in Section \ref{sec:4_DS-PPO}. Subsequently, simulation results are provided and discussed in Section \ref{sec:5_simulation_results}. Finally, the paper is concluded in Section \ref{sec:6_conclusion}.

\textit{Notations:} In this paper, $a$, $\boldsymbol{a}$, $\boldsymbol{A}$, and $\mathcal{A}$ denote the scalars, the vectors, the matrices, and the defined set, respectively. Specially, $\boldsymbol 0$ and $\boldsymbol 1$ represent the all-zero vector and the all-one vector, respectively. The superscript ${\rm T}$ represents the transpose operation. $\mathbb{R}^{x \times y}$ and $\{a_i\}$ denote the set of $x \times y$ real-valued matrix and the set of $a_i$, respectively. Additionally, $\|\cdot\|_{\rm F}$, $\|\cdot\|$, $|\cdot|$, $\lfloor\cdot\rfloor$, $\mathbb{E}[\cdot]$, $\nabla$, and $\odot$ denote the Frobenius norm operator, the Euclidean norm operation, the absolute value operation, the floor function, the expectation operator, the gradient operator, and the Hadamard product operator, respectively. $\mathcal{N}(\mu,\sigma^2)$ represents the distribution of a real-valued Gaussian noise with mean $\mu$ and variance $\sigma^2$. Moreover, $[\cdot]^+=\max\{\cdot,0\}$ and ${\rm mod}(\cdot)$ are the positive function and the modulus operator, respectively.

\section{System Model}
\label{sec:2_system_model}
\subsection{VLC Network}
\begin{figure*}[t!]
\centering
\includegraphics[width=0.9\textwidth]{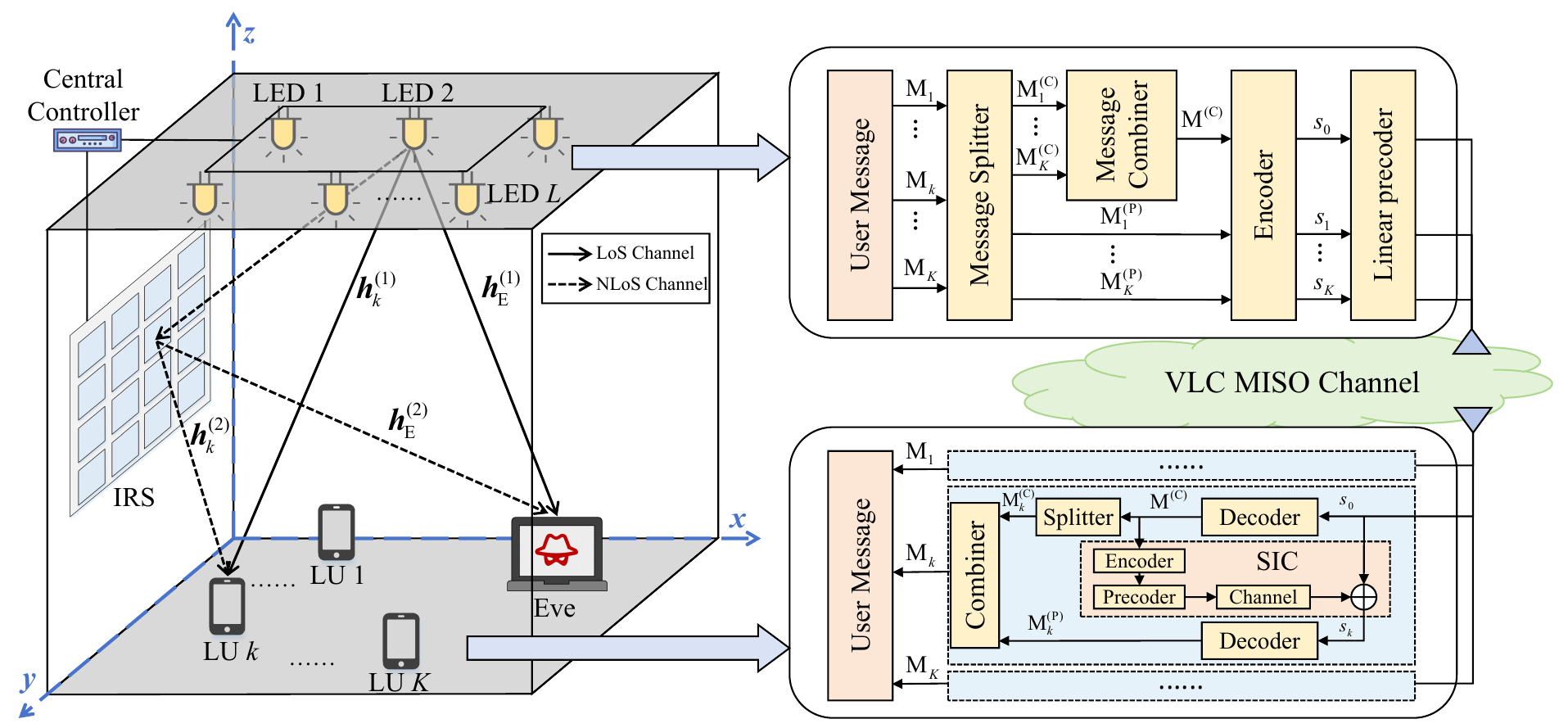}
\caption{{{Illustration of the IRS-assisted VLC network with RSMA. The IRS reflects LED signals to enhance communication with user devices (i.e., all LUs) while mitigating interference from an Eve. RSMA is employed to split and encode user messages, improving data transmission and interference management across the VLC MISO channel.}}}
\label{fig_model}
\end{figure*}

Fig.~\ref{fig_model} illustrates the considered indoor IRS-assisted VLC network with RSMA. In this network, an IRS, equipped with $N$ passive mirror-like reflecting elements, is mounted on the wall\cite{1004299}. With the assistance of the IRS, $L$ LEDs provide high-quality communication services to $K$ LUs. All LEDs and the IRS are interconnected by wires and collectively controlled by a central controller. Simultaneously, an Eve in this network attempts to intercept the information sent to the $K$ LUs through both LoS and non-line-of-sight (NLoS) channels. Besides, each LU is equipped with a photo-detector (PD) to receive the modulated downlink VLC signals. Without loss of generality, let $l$ and $k$ denote the $l$-th LED and the $k$-th LU, respectively, $\forall l\in \mathcal{L}\triangleq \{1,2,3,...,L\}$, $\forall k\in \mathcal{K}\triangleq \{1,2,3,...,K\}$.

As shown in Fig.~\ref{fig_model}, let $M_k$ denote the information intended for the $k$-th LU, with the RSMA scheme, $M_k$ is first divided into two parts, one is the common part $M_k^{\rm (C)}$ and the other one is the private part $M_k^{\rm (P)}$. Subsequently, the common parts of all LUs are combined into $M^{\rm (C)}$, and encoded into a real-valued common stream\footnote{Intensity modulation and direct detection (IM/DD) scheme is typically adopted for VLC systems, where electrical signals are converted to real-valued and non-negative waveforms to drive the LEDs. Thus, the transmit signals of VLC are real values\cite{10552701,1004299}.}, $s_0 \in \mathbb{R}$. While the private part of the $k$-th LU is separately encoded into a private stream $s_k \in \mathbb{R}$\cite{10301507}. Also, $s_i$ is normalized with unit variance and zero mean, i.e., $\mathbb{E}[|s_i|^2]=1$, $i\in\{0\}\cup\mathcal K$. The transmitted VLC signal, denoted as $\boldsymbol x=[x_1,\ldots,x_l,\ldots,x_L]^{\rm T}\in \mathbb{R}^{L\times 1}$, is
\begin{equation}\label{eq_signal}
\boldsymbol{x} = \underbrace{\boldsymbol{v}_0 s_0}_{\text{Common part}} + \underbrace{{\sum}_{k=1}^K \boldsymbol{v}_k s_k}_{\text{Private part}}+\boldsymbol{d}_{\rm {DC}},
\end{equation}
where $ \boldsymbol{v}_0=[v_{0,1},...,v_{0,L}]^{\rm T}$ and $ \boldsymbol{v}_k=[v_{k,1},...,v_{k,L}]^{\rm T}$ denote the beamforming vector for $s_0$ and $s_k$, respectively.  $\boldsymbol d_{\rm DC} = [I_{\rm DC},\ldots, I_{\rm DC}]^T \in \mathbb{R}^{L\times 1}$, where $I_{\rm DC}$ is the direct current of each LED. Moreover, practical LEDs operate within a limited linear operating region to avoid clipping\cite{9493176}, the amplitude of $x_l$ is constrained by $I_{\min}$ and $I_{\max}$, i.e.,
\begin{equation}\label{eq_transmit_limit}
{\sum}_{i=0}^K{|v_{i,l}|}\leq\Delta,\forall l \in \mathcal{L},
\end{equation}
where $\Delta=\min(I_{\rm {DC}}-I_{\rm {min}},I_{\rm {max}}-I_{\rm {DC}})$. $I_{\rm {min}}$ and $I_{\rm {max}}$ represents the minimum and maximum driving current, respectively. Also, the total electrical power consumed by the considered VLC network can be expressed as
\begin{equation}\label{eq_total_power}
\begin{split}
P_{\rm T}= {\sum}_{i=0}^K || \boldsymbol{v}_i ||^2 +U_{\rm {LED}}\mathbf{1}_L^{\rm T}\boldsymbol{d}_{\rm {DC}}+ P_{\rm C},
\end{split}
\end{equation}
where $\mathbf{1}_L\in \mathbb{R}^{L\times 1}$ is the all-one vector and $U_{\rm {LED}}$ denotes the forward voltage of the LED\cite{9463422}. $P_{\rm C} \geq 0$ represents the fixed power consumption of the circuit module and IRS of the considered system.
\subsection{VLC Channel Model}
The channels of VLC system are classified into two types, i.e., LoS channel and NLoS channel. Without loss of generality, {we consider} that the VLC channel state information (CSI) of both two types of channels associated with all LUs is known at the central controller, which can be achieved through various existing channel estimation methods\cite{9714890,10299678,9756553}.
\subsubsection{LoS Channels}
Since LED transmitters typically adhere the Lambertian model, the VLC channel gain can be determined based on the geometric position of the LED transmitters and the LU\cite{10024150}. The channel gain between the $l$-th LED and the $k$-th LU can be expressed as
\begin{align}\label{eq_LoS_gains_user}
 h_{l,k}^{(1)} &=\frac{(m + 1) A_{\rm P}}{2 \pi d^2_{l,k}}\cos^{m} (\phi_{l,k}) \cos(\psi_{l,k})g_{\rm of}f(\psi_{l,k}),\forall l,
\end{align}
where $m = - {1}/{\rm log_2(\cos(\Omega_{1/2}))}$ represents the Lambertian index. $A_{\rm P}$ denotes the physical area of the PD. $\Omega_{1/2}$, $\phi_{l,k}$, and $\psi_{l,k}$ represent the half intensity view angle, the irradiance angle and the incidence angle, respectively. $d_{l,k}$ is the LoS transmission distance between the $l$-th LED and the $k$-th LU, and $g_{\rm of}\in \mathbb{R}$ denotes the optical filter gain. Besides, $f(\psi_{l,k})$ represents the optical concentrator gain\cite{10684100,10024150,7106482}, i.e.,
\begin{equation}\label{eq_concentrator_gain}
f(\psi_{l,k})=
\begin{cases}
\frac{\kappa^2}{\sin^2(\Psi_{\rm F})}, & \mbox{if }0\leq \psi_{l,k}\leq \Psi_{\rm F},\\
0, & \mbox{otherwise},
\end{cases}
\end{equation}
where $\kappa$ is the refractive index and $\Psi_{\rm F}$ stands for the PD's fields of view (FoV). {Following \cite{9834293,10404069,10128157}, Eve's CSI is assumed to be known at the central controller,} thus the channel gain between the $l$-th LED and Eve is
\begin{align}\label{eq_LoS_gains_Eve}
 h_{l,\rm E}^{(1)} &=\frac{(m + 1) A_{\rm P}}{2 \pi d^2_{l,\rm E}}\cos^{m} (\phi_{l,\rm E}) \cos(\psi_{l,\rm E})g_{\rm of}f(\psi_{l,\rm E}),\forall l,
\end{align}
where $\phi_{l,\rm E}$ and $\psi_{l,\rm E}$ {are} the irradiance angle and the incidence angle, respectively. $d_{l,\rm E}$ denotes the LoS transmission distance between the $l$-th LED and the Eve.


Thus, the received signals at the $k$-th LU and Eve via the LoS channels can be expressed as
\begin{align}\label{eq_Los_receives}
\left\{
  \begin{aligned}
  y_k^{(1)}&=\boldsymbol{h}_k^{(1){\rm T}}(\boldsymbol{v}_0 s_0 + {\sum}_{k=1}^K \boldsymbol{v}_k s_k+\boldsymbol{d}_{\rm {DC}}),\\
  y_{\rm E}^{(1)}&=\boldsymbol{h}_{\rm E}^{(1){\rm T}}(\boldsymbol{v}_0 s_0 + {\sum}_{k=1}^K \boldsymbol{v}_k s_k+\boldsymbol{d}_{\rm {DC}}),
  \end{aligned}
\right.
\end{align}
respectively, where $\boldsymbol{h}^{(1)}_k=[h_{1,k}^{(1)},\ldots,h_{L,k}^{(1)}]^{\rm T}$, $\boldsymbol{h}_{\rm E}^{(1)}=[h_{1,\rm E}^{(1)},\ldots,h_{L,\rm E}^{(1)}]^{\rm T}$.
\subsubsection{NLoS Channels}
Optical reflections are categorized into diffuse reflection and specular reflection, where the former is always neglected due to its much lower channel gain than that of the latter\cite{7106482,10024150}. Thus, the NLoS channel gain from the $l$-th LED to the $k$-th LU through the $n$-th reflecting element can be modeled as
\begin{align}\label{eq_NLoS_gains_user}
\begin{aligned}
&g_{l,n,k}\\
&=\frac{\rho(m + 1) A_{\rm P}}{2 \pi (d_{l,n}+d_{n,k})^2}\cos^{m} (\phi_{l,n}) \cos(\psi_{n,k})g_{\rm of}f(\psi_{n,k}),
\end{aligned}
\end{align}
where $n \in \mathcal N \triangleq \{1,2,\ldots,N\}$ and $\rho$ represents the reflection coefficient of the IRS. $d_{l,n}$ and $d_{n,k}$ are the distances from the $l$-th LED to the $n$-th IRS element and that from the $n$-th element to the $k$-th LU, respectively.

Since optical frequency beams are highly concentrated along a straight line, the NLoS channel via an IRS element is only effective when the element is aligned with an LED-LU pair\cite{10024150}. Therefore, the achievable NLoS channel is determined by the channel gain $\boldsymbol g_{l,k}\triangleq[g_{l,1,k},\ldots,g_{l,N,k}]^{\rm T}$ and two binary alignment matrices, i.e.,
\begin{align}\label{eq_alignment_matrices}
\begin{aligned}
\boldsymbol A&=\left[
\begin{matrix}
  a_{1,1} & \cdots & a_{1,K} \\
  \vdots & \ddots & \vdots \\
  a_{N,1} & \cdots & a_{N,K} 
\end{matrix}\right],
\boldsymbol B&=\left[
\begin{matrix}
  b_{1,1} & \cdots & b_{1,L} \\
  \vdots & \ddots & \vdots \\
  b_{N,1} & \cdots & b_{N,L} 
\end{matrix}\right],
\end{aligned}
\end{align}
where $\boldsymbol A$ and $\boldsymbol B$ represent the alignment matrices between the IRS elements and LUs and that between LEDs and IRS elements, respectively. Particularly, $a_{n,k}=1$ or $b_{n,l}=1$ indicates an alignment, while $a_{n,k}=0$ or $b_{n,l}=0$ indicates misalignment. Hence, the NLoS channel between the $l$-th LED and the $k$-th LU can be expressed as
\begin{equation}\label{eq_nlos_channel_item}
h_{l,k}^{(2)}={\sum}_{n=1}^{N}a_{n,k}b_{n,l}g_{l,n,k}=(\boldsymbol a_k\odot\boldsymbol b_l)^{\rm T}\boldsymbol g_{l,k},
\end{equation}
where $\boldsymbol a_k$ and $\boldsymbol b_l$ are the $k$-th column of $\boldsymbol A$ and the $l$-th column of $\boldsymbol B$, respectively.

Furthermore, since the reflected beam path is almost straight, each IRS element can be assigned to a maximum of an LED and an LU (an LED-LU pair)\cite{9526581}, i.e., 
\begin{align}\label{eq_alignment_contraints}
\left\{
\begin{aligned}
&{\sum}_{k=1}^K a_{n,k}\leq 1,~\forall n \in \mathcal{N},\\
&{\sum}_{l=1}^L b_{n,l}\leq 1,~\forall n \in \mathcal{N}.
\end{aligned}
\right.
\end{align}
Therefore, let $\boldsymbol h^{(2)}_k = [h_{1,k}^{(2)},\ldots,h_{L,k}^{(2)}]^{\rm T}$ denote the NLoS channel gain of the $k$-th LU, the received signals at the $k$-th LU over the NLoS channel can be given by
\begin{align}\label{eq_NLoS_receives}
\begin{aligned}
y_k^{(2)}&=\boldsymbol h_k^{(2){\rm T}}(\boldsymbol{v}_0 s_0 + {\sum}_{k=1}^K \boldsymbol{v}_k s_k+\boldsymbol{I}_{\rm {DC}}).
\end{aligned}
\end{align}

To enhance the information security and degrade the achieved SINR at Eve, IRS elements would not be aligned with Eve, i.e., $\boldsymbol h_{\rm E}^{(2)}=[h_{1,{\rm E}}^{(2)},\ldots,h_{L,{\rm E}}^{(2)}]^{\rm T}=\boldsymbol 0$.
\subsubsection{IRS-Assisted Channel}
Combining the LoS path with the NLoS path, the channel gain of the $k$-th LU is $\boldsymbol h_k=\boldsymbol h_k^{(1)}+\boldsymbol h_k^{(2)}=\boldsymbol h_k^{(1)}+\sum_{l=1}^{L} (\boldsymbol a_k \odot \boldsymbol b_l)^{\rm T} \boldsymbol g_{l,k}$, based on which, the received signals of the $k$-th LU and Eve can be expressed as
\begin{align}\label{eq_total_receive}
\left\{
\begin{aligned}
y_k&=y_k^{(1)}+y_k^{(2)}+n_k\\
&=\boldsymbol h_k^{\rm T}(\boldsymbol{v}_0 s_0 + {\sum}_{k=1}^K \boldsymbol{v}_k s_k+\boldsymbol{d}_{\rm DC})+{n_k},\\
y_{\rm E}&=y_{\rm E}^{(1)}+n_{\rm E}\\
&=\boldsymbol h_{\rm E}^{(1){\rm T}}(\boldsymbol{v}_0 s_0 + {\sum}_{k=1}^K \boldsymbol{v}_k s_k+\boldsymbol{d}_{\rm DC})+n_{\rm E},
\end{aligned}
\right.
\end{align}
respectively, where $n_k \sim\mathcal{N}\left(0,\sigma_k^2\right)$ and $n_{E} \sim\mathcal{N}\left(0,\sigma_{\rm E}^2\right)$ {are} additive white Gaussian noise at the $k$-th LU and the Eve, respectively. $\sigma_k^2$ and $\sigma_{\rm E}^2$ are the corresponding noise power density.
\section{Problem Formulation}
\label{sec:3_problem_formulation}
In this section, we analyze the process of information transmission, followed by the formulation of the optimization problem.
\subsection{Information Transmission}
\begin{figure*}[!htb]
\centering
\begin{align}
R_k^{(\rm C)}=&~
\begin{aligned}\label{eq_receive_common}
\log_2\left(1+\frac{e}{2\pi}\frac{|\boldsymbol{h}_{k}^{\rm T} \boldsymbol{v}_0|^2}{{\sum\limits}_{i=1}^K {|{\boldsymbol{h}_{k}^{\rm T}}\boldsymbol{v}_i|^2+\sigma_k^2}}\right),~\forall k \in \mathcal{K},
\end{aligned}\\
R_k^{(\rm P)}=&~
\begin{aligned}\label{eq_receive_private}
\log_2\left(1+\frac{e}{2\pi}\frac{|\boldsymbol{h}_{k}^{\rm T} \boldsymbol{v}_k|^2}{{\sum\limits}_{i=1,i \ne k}^K {|{\boldsymbol{h}_{k}^{\rm T}}\boldsymbol{v}_i|^2+\sigma_k^2}}\right),~\forall k \in \mathcal{K}.
\end{aligned}
\end{align}
\hrulefill
\end{figure*}
With the RSMA scheme, the achieved information rates at the $k$-th LU by decoding common stream and private streams, denoted by $R_k^{(\rm C)}$ and $R_k^{(\rm P)}$, can be expressed {as in} (\ref{eq_receive_common}) and (\ref{eq_receive_private}), respectively. To ensure successful decoding of the common stream by all LUs, the received common information at all LUs should satisfy ${\sum}_{k=1}^{K} c_k \leq \min \limits_{k}\{R_1^{(\rm C)},\ldots,R_k^{(\rm C)},\ldots,R_K^{(\rm C)}\}$, where $c_k$ represents the common information rate of the $k$-th LU\cite{9831440}. Therefore, the total information data rate at the $k$-th LU is
\begin{equation}\label{eq_total_receive_rate}
R_k^{\rm {total}}=c_k+R_k^{(\rm P)},\forall k.
\end{equation}

On the other hand, when Eve attempts to eavesdrop the data stream delivered to the LUs, the common eavesdropping rate and private eavesdropping rate, denoted by $R_{\rm E}^{(\rm C)}$ and $R_{{\rm E},k}^{(\rm P)}$, can be expressed {as in} (\ref{eq_eve_common_rate}) and (\ref{eq_eve_private_rate}), respectively. Therefore, the total secrecy rate is given by
\begin{figure*}[!htb]
\centering
\begin{align}
R_{\rm E}^{(\rm C)}=&~
\begin{aligned}\label{eq_eve_common_rate}
\log_2\left(1+\frac{e}{2\pi}\frac{|\boldsymbol{h}_{\rm E}^{\rm (1)T} \boldsymbol{v}_0|^2}{{\sum\limits}_{i=1}^K {|{\boldsymbol{h}_{\rm E}^{\rm (1)T}}\boldsymbol{v}_i|^2+\sigma_{\rm E}^2}}\right),
\end{aligned}\\
R_{{\rm E},k}^{(\rm P)}=&~
\begin{aligned}\label{eq_eve_private_rate}
\log_2\left(1+\frac{e}{2\pi}\frac{|\boldsymbol{h}_{\rm E}^{\rm (1)T} \boldsymbol{v}_k|^2}{|\boldsymbol{h}_{\rm E}^{\rm (1)T} \boldsymbol{v}_0|^2+{\sum\limits}_{i=1,i \ne k}^K {|{\boldsymbol{h}_{\rm E}^{\rm (1)T}}\boldsymbol{v}_i|^2+\sigma_{\rm E}^2}}\right),~\forall k \in \mathcal{K}.
\end{aligned}
\end{align}
\hrulefill
\end{figure*}

\begin{align}\label{eq_secrecy_rate}
\begin{aligned}
\hat R_{\rm T}&=\hat R^{(\rm C)}+\hat R^{(\rm P)}\\
&=\left[{\sum}_{k=1}^{K} c_k-R_{\rm E}^{\rm (C)}\right]^++{\sum}_{k=1}^{K} \left[R_k^{(\rm P)}-R_{{\rm E},k}^{(\rm P)}\right]^+,
\end{aligned}
\end{align}
where $\hat R^{(\rm C)}$ and $\hat R^{(\rm P)}$ denote the total secrecy rate of common stream and that of private streams\cite{10078728}, respectively.

As SEE is defined as the ratio of the total secrecy rate to the total power consumption, the SEE of the considered system is given by
\begin{equation}\label{eq_SEE}
{\rm {SEE}}=\frac{\hat R_{\rm T}}{P_{\rm T}}=\frac{{\sum}_{k=1}^{K}\hat R_k}{{\sum}_{n=0}^K || \boldsymbol{v}_n ||^2 +U_{\rm LED}\mathbf{1}_L^{\rm T}\boldsymbol{d}_{\rm {DC}}+ P_{\rm C}}.
\end{equation}
\subsection{Optimization Problem}
Our goal is to maximize the SEE of the considered system by jointly optimizing the beamforming vectors, the DC bias, the common information rate, and the alignment matrices of the IRS elements. {Meanwhile,} constraints of the transmit power budget, QoS requirements of each LU, the common information rate, and the linear operation region of LEDs {should be satisfied}. Accordingly, from the mathematical perspective, {we formulate the} optimization problem as
\begin{subequations}
\begin{align}
\boldsymbol{\mathcal P_0}:& \mathop {\max} \limits_{\{\{\boldsymbol{v}_i,c_k\},\boldsymbol{d}_{\rm {DC}},\boldsymbol{A},\boldsymbol{B}\}}{\rm {SEE}}({\{\boldsymbol{v}_i,
c_k\},\boldsymbol{d}_{\rm {DC}},\boldsymbol{A},\boldsymbol{B}})
\label{eq_P0}\\
{\rm s.t.}~
&~ c_k+R_k^{(\rm P)} \ge \zeta_k, ~ \forall k \in \mathcal K,
\label{P0_C1}\\
&~ {\sum}_{k=1}^K c_k \leq \mathop {\min} \limits_{k} ~ \{R^{\rm (C)}_1,R^{\rm (C)}_2,...,R^{\rm (C)}_K\},
\label{P0_C2}\\
&~ P_{\rm {T}} \leq P_{\rm{max}},
\label{P0_C3}\\
&~ {\sum}_{i=0}^K{|v_{i,l}|}\leq\Delta,~\forall l \in \mathcal L,
\label{P0_C4}\\
&~ {\sum}_{k=1}^K a_{n,k}\leq 1,~\forall n \in \mathcal{N},
\label{P0_C5}\\
&~ {\sum}_{l=1}^L b_{n,l}\leq 1,~\forall n \in \mathcal{N},
\label{P0_C6}\\
&~ a_{n,k},b_{n,l} \in \{0,1\},~\forall l \in \mathcal L,\forall k \in \mathcal K,
\label{P0_C7}\\
&~ c_k \geq 0, ~ \forall k \in \mathcal K,
\label{P0_C8}\\
&~ \Delta \geq 0.
\label{P0_C9}
\end{align}
\end{subequations}
In problem $\boldsymbol{\mathcal P_0}$, constraint (\ref{P0_C1}) ensures that the total received information rate of the $k$-th LU should not be less than the QoS threshold $\zeta_k$. Constraints (\ref{P0_C2}) and (\ref{P0_C8}) are imposed by the RSMA scheme on the common information rate to ensure that the common stream is successfully decoded by each LU. Also, constraint (\ref{P0_C3}) guarantees that the total power consumption of this system does not exceed the maximum power budget $P_{\rm max}$. 
Furthermore, constraints (\ref{P0_C4}) and (\ref{P0_C9}) are invoked to ensure that each LED operates within its linear operating region. Constraints (\ref{P0_C5}), (\ref{P0_C6}), and (\ref{P0_C7}) are due to the definitions of $\boldsymbol A$ and $\boldsymbol B$.
\section{Proposed DRL-based Approach}
\label{sec:4_DS-PPO}
Since $\boldsymbol{\mathcal P_0}$ involves the mixed variables, the non-convex objective function and the constraints in (\ref{eq_P0})-(\ref{P0_C3}), its globally optimal solution is difficult to {obtain using} available polynomial-time algorithms. To address this challenge, we first relax $\boldsymbol{\mathcal P_0}$ and reformulate it as an MDP, based on which, a DS-PPO approach is proposed.

\subsection{Optimization Problem Reformulation}
Due to constraints (\ref{P0_C5})-(\ref{P0_C7}), the columns of $\boldsymbol A$ as well as that of $\boldsymbol B$ are orthogonal to each other. Therefore, according to \cite{10024150}, the problem $\boldsymbol{\mathcal P_0}$ can be equivalently transformed into $\boldsymbol{\mathcal P_1}$ as follows:
\begin{subequations}
\begin{align}
\boldsymbol{\mathcal P_1}:& \mathop {\max} \limits_{\{\{\boldsymbol{v}_i,c_k\},\boldsymbol{d}_{\rm {DC}},\boldsymbol{Q}\}}{\rm {SEE}}({\{\boldsymbol{v}_i,c_k\},\boldsymbol{d}_{\rm {DC}},\boldsymbol{Q}})
\label{eq_P1}\\
{\rm s.t.}~
&~ {\sum}_{p=1}^{LK} q_{n,p}\leq 1,~\forall n \in \mathcal{N},
\label{P1_C1}\\
&~ q_{n,p} \in \{0,1\},~\forall p \in \mathcal P,
\label{P1_C2}\\
&~ \mathrm {(\ref{P0_C1}), (\ref{P0_C2}), (\ref{P0_C3}), (\ref{P0_C4}), (\ref{P0_C8}), (\ref{P0_C9})},
\label{P1_C3}
\end{align}
\end{subequations}
where $\mathcal P\triangleq\{1,2,\ldots,LK\}$ is the index set for the merged matrix $\boldsymbol Q \triangleq [\boldsymbol q_1,\ldots,\boldsymbol q_{LK}]\in \{0,1\}^{N\times LK}$, which is a matrix with columns defined as
\begin{equation}\label{eq_alignment_q}
\boldsymbol q_{k+(l-1)K}=\boldsymbol a_k\odot \boldsymbol b_l,~\forall k \in \mathcal K,~\forall l \in \mathcal L,
\end{equation}
where $q_{n,k+(l-1)K}=1$ indicates that the reflected channel between the $l$-th LED and $k$-th LU via the $n$-th IRS element is established.

Due to the existence of integer variables, it remains challenging in solving problem $\boldsymbol{\mathcal P_1}$. To tackle this issue, a feasible approach is to first optimize the relaxed form of $\boldsymbol{\mathcal P_1}$, after which the binary results could be derived by the minimum distance projection criterion\cite{10433073}. As such, by converting the binary matrix Q into a continuous one, $\boldsymbol{\mathcal P_1}$ can be relaxed as
\begin{subequations}
\begin{align}
\boldsymbol{\mathcal P_2}:& \mathop {\max} \limits_{\{\{\boldsymbol{v}_i,c_k\},\boldsymbol{d}_{\rm {DC}},\boldsymbol{\widetilde Q}\}}{\rm {SEE}}({\{\boldsymbol{v}_i,
c_k\},\boldsymbol{d}_{\rm {DC}},\boldsymbol{\widetilde Q}})
\label{eq_P2}\\
{\rm s.t.}~
&~ 0 \leq \widetilde{q}_{n,p}\leq 1,~\forall n \in \mathcal{N},~\forall p \in \mathcal P,
\label{P2_C1}\\
&~ \mathrm {(\ref{P0_C1}), (\ref{P0_C2}), (\ref{P0_C3}), (\ref{P0_C4}), (\ref{P0_C8}), (\ref{P0_C9})},
\label{P2_C2}
\end{align}
\end{subequations}
where $\boldsymbol{\widetilde Q}$ is the relaxed form of $\boldsymbol Q$. With the minimum distance projection criterion, the relaxed results of $\boldsymbol{\mathcal P_2}$ can be transformed into binary results. Particularly, the optimal projection index $\bar p$ for the $n$-th IRS element can be computed as $\bar p=\arg \mathop {\min}_p ||\widetilde{\boldsymbol q}_n-\boldsymbol t_p||^2$, where $\widetilde{\boldsymbol q}_n$ is the $n$-th row of $\widetilde{\boldsymbol Q}$, $\boldsymbol t_p$ denotes an auxiliary vector with the $p$-th element being 1. Moreover, $\boldsymbol t_0$ is defined as a zero vector. Thus, the indices of the corresponding LED-LU pair is given by
\begin{align}\label{eq_indexes}
\left\{
\begin{aligned}
l&=\left\lfloor \frac{\bar p-1}{N}\right\rfloor+1,\\
k&={\rm mod} ({\bar p-1},N)+1.
\end{aligned}
\right.
\end{align}

\subsection{MDP Formulation}
{An MDP} consists of an agent, {i.e., the transmitter,} and 4-tuples $<\mathcal A,\mathcal S,\mathcal R, \gamma>$, with $\mathcal A,\mathcal S,\mathcal R$, and $\gamma$ denoting the action space, state space, reward function, and discount factor, respectively\cite{MDP}. According to problem $\boldsymbol{\mathcal P_2}$, the action space, the state space, and the reward function are defined as follows.
\subsubsection{Action Space}
The action space is composed of the beamforming vectors $\{\boldsymbol{v}_k\}$, direct current bias $\boldsymbol d_{\rm DC}$, common information rate $\{c_k\}$, and the relaxed IRS alignment matrix $\boldsymbol{\widetilde Q}$, i.e., $\mathcal A \triangleq \{\{\boldsymbol{v}_k\},\{ c_k\},\boldsymbol d_{\rm DC},\boldsymbol{\widetilde Q}\} $.

Due to the complexity of neural networks in handling the beamforming vectors, $ \boldsymbol{v}_k$ is divided into two parts for separate processing\cite{10032267}, i.e., 
\begin{align}
\boldsymbol{v}_k = ||\boldsymbol{v}_k|| \cdot \overline{\boldsymbol{v}}_k,
\end{align}
where $||\boldsymbol{v}_k||$ and $\overline{\boldsymbol{v}}_k$ represent the transmit power allocated to the $k$-th stream $s_k$ and the normalized direction of $\boldsymbol{v}_k$, respectively. Particularly, the transmit power $\|\boldsymbol{v}_k\|$ is first obtained by the hyperbolic tangent function, and then scaled by $\sqrt{P_{\rm {max}}}$ to satisfy the constraint of total power budget \cite{10107766}, i.e.,
\begin{equation}\label{eq_vk_power}
||\boldsymbol{v}_k||=\frac{\sqrt{P_{\rm {max}}}}{2}\left(\frac{e^{o_k^{\rm {beam}}}-e^{-o_k^{\rm {beam}}}}{e^{o_k^{\rm {beam}}}+e^{-o_k^{\rm {beam}}}}+1 \right)
\end{equation}
where $\frac{e^{o_k^{\rm {beam}}}-e^{-o_k^{\rm {beam}}}}{e^{o_k^{\rm {beam}}}+e^{-o_k^{\rm {beam}}}} \in [-1,1]$ is the hyperbolic tangent function, serving as an activation function to ensure that the outputs of neural networks satisfies the corresponding constraints\cite{10032267}. $o_k^{\rm {beam}}=\digamma(\omega o_k^{\rm {beam'}}+\beta)$ is the output of the activation function in terms of the selected transmit power, where $\digamma(\cdot)$, $o_k^{\rm {beam'}}$, $\omega$, and $\beta$ denote the activation function, the output of the previous layer, the corresponding weight, and the corresponding bias, respectively.

To simplify the action space of DS-PPO, MRT and ZF schemes are adopted to design beamforming vectors. Since the common stream is decoded by all LUs, and it does not cause interference to the private streams, MRT is exploited to design the beamforming associated with the common stream. On the other hand, each private stream is only decoded by its corresponding LU, based on which, ZF is adopted to determined the beamforming vectors associated with private streams, aiming to effectively mitigate the IUI\cite{10032267}. Specifically, the direction of beamforming vectors determined by ZF scheme can be expressed as
\begin{equation}\label{eq_vk_direction}
\overline{\boldsymbol v}_k=\frac{\boldsymbol{H}^{-1}}{||\boldsymbol{H}^{-1}||_{\rm F}}=\frac{\boldsymbol{H}^\dag+\boldsymbol{PR}}
{||\boldsymbol{H}^\dag+\boldsymbol{PR}||_{\rm F}},
\end{equation}
where $\boldsymbol{H}^\dag=\boldsymbol{H}^{\rm T}(\boldsymbol{H}\boldsymbol{H}^{\rm T})^{-1}$ is the pseudo-inverse of the channel matrix $\boldsymbol{H}$, $\boldsymbol P$ is the orthogonal projection of $\boldsymbol H$ onto its null space, and $\boldsymbol R$ is an arbitrary matrix. Without loss of generality, let $\boldsymbol R$ be the all-zero matrix, then $\overline{\boldsymbol{v}}_k = {\boldsymbol H^\dag}/{\|\boldsymbol H^\dag\|_{\rm F}}$.

To sum up, the direction of beamforming vectors designed by MRT and ZF scheme is given by
\begin{equation}\label{eq_MRT_ZF}
\overline{\boldsymbol v}_k=
\begin{cases}
 \overbrace{\frac{\sum_{i=0}^{K}{\boldsymbol{h}}_i}{||\sum_{i=0}^{K}{\boldsymbol{h}}_i||_{\rm F}}}^{\text{MRT for the common stream}} &,k=0, \\ \\
  \underbrace{\frac{\boldsymbol{H}^T(\boldsymbol{H}\boldsymbol{H}^T)^{-1}}
  {||\boldsymbol{H}^T(\boldsymbol{H}\boldsymbol{H}^T)^{-1}||_{\rm F}}}_{\text{ZF for the private stream}} &,k \neq 0,
\end{cases}
\end{equation}
where $\boldsymbol H \triangleq [{\boldsymbol{h}}_1,{\boldsymbol{h}}_2,\ldots,{\boldsymbol{h}}_K] $.

Moreover, $I_{\rm {DC}}$ and $c_k$ are first calculated by the hyperbolic tangent function, and then scaled by $I_{\rm {max}}$ and $\mathop {\min} \limits_{k} ~ \{R^{\rm (C)}_k\}$, respectively, to satisfy the corresponding constraints, i.e.,
\begin{align}
I_{\rm {DC}}&=
\begin{aligned}\label{eq_I_DC}
\frac{I_{\rm {max}}}{2}\left(\frac{e^{o_k^{\rm {dc}}}-e^{-o_k^{\rm {dc}}}}
{e^{o_k^{\rm {dc}}}+e^{-o_k^{\rm {dc}}}}+1 \right),
\end{aligned}\\
c_k&=
\begin{aligned}\label{eq_ck}
\frac{\mathop {\min} \limits_{k} ~ \{R^{\rm (C)}_k\}}{2}\left(\frac{e^{o_k^{\rm com}}-e^{-o_k^{\rm com}}}{e^{o_k^{\rm com}}+e^{-o_k^{\rm com}}}+1 \right),
\end{aligned}
\end{align}
where $o_k^{\rm {dc}}$ and $ o_k^{\rm com}$ are the corresponding outputs of the activation function.

Thus, the dimension of $\mathcal{A}$ is
\begin{equation}\label{eq_action_shape}
(\underbrace{K + 1}_{\boldsymbol{v}} +\underbrace{L}_{\boldsymbol d_{\rm {DC}}}+\underbrace{NLK}_{\boldsymbol{\widetilde Q}}+\underbrace{K}_{c_k}).
\end{equation}
\subsubsection{State Space}
The state space should contain as much environmental information relevant to the optimization problem as possible\cite{10032267}. In the considered system, the state space includes the action $\boldsymbol{a}$, reward $r$, and SINR $\boldsymbol{u}$, i.e.,
\begin{equation}\label{eq_action_space}
\begin{aligned}
\mathcal{S}&\triangleq\{\boldsymbol{a},\boldsymbol{u},r\}\\
&\triangleq\{\boldsymbol a,\underbrace{\{\gamma^{\rm C}_{\rm {LU}}\},\{ \gamma^{\rm P}_{\rm {LU}}\},\{\gamma^{\rm C}_{\rm {Eve}}\},\{ \gamma^{\rm P}_{\rm {Eve}}\}}_{\boldsymbol{u}},r\},
\end{aligned}
\end{equation}
where $\boldsymbol{a} \in \mathcal{A}$ represents the action chosen at the previous step. $\gamma^{\rm C}_{\rm {LU}}$ and $\gamma^{\rm P}_{\rm {LU}}$ denote the SINR at LUs by decoding the common stream and private streams, respectively. Moreover, $\gamma^{\rm C}_{\rm {Eve}}$ and $\gamma^{\rm P}_{\rm {Eve}}$ denote the SINR at the Eve by eavesdropping the common stream and private streams, respectively. The reward $r$ is calculated by the state-action pair, which can intuitively reflect the ability of the agent in addressing the problem $\boldsymbol{\mathcal P_1}$.

As a result, the dimension of state space $\mathcal S$ is
\begin{equation}\label{eq_state_shape}
(\underbrace{NLK+2K+L+1}_{\boldsymbol{a}}+\underbrace{3K+1}_{ \boldsymbol{u}}+\underbrace{1}_{r}).
\end{equation}
\subsubsection{Reward Function}
In our work, the reward is calculated based on the SEE and the penalties associated with the constraints in problem $\boldsymbol{\mathcal{P}_1}$. As such, to maximize the SEE while satisfying the constraints, the reward function is defined as
\begin{equation}\label{eq_reward}
\begin{aligned}
r=&\overbrace{{\rm {SEE}}(\{\boldsymbol v_k,c_k\},\boldsymbol d_{\rm {DC}},\boldsymbol \Phi)}^{\rm{Goal}}\\
&\quad\times \underbrace{\Gamma_{\rm {power}}\times\Gamma_{\rm {common}}\times\Gamma_{\rm {QoS}}\times \Gamma_{\rm {linear}}}_{\rm{Penalty~terms}},
\end{aligned}
\end{equation}
where $\Gamma_{\rm {power}}$, $\Gamma_{\rm {common}}$, $\Gamma_{\rm {QoS}}$, and $\Gamma_{\rm {linear}}$ are the penalties for actions associated with constraints (\ref{P0_C3}), (\ref{P0_C2}), (\ref{P0_C1}), and (\ref{P0_C4}), respectively, and can be given by
\begin{align}
\Gamma_{\rm {power}}&=\left\{
\begin{aligned}\label{eq_constraint_power}
1, &~~P_{\rm T}(\{\boldsymbol{v}_k\},\boldsymbol{d}_{\rm {DC}})\leq P_{\max}, \\
0, &~~P_{\rm T}(\{\boldsymbol{v}_k\},\boldsymbol{d}_{\rm {DC}})> P_{\max},
\end{aligned}\right.\\
\Gamma_{\rm {common}}&=\left\{
\begin{aligned}\label{eq_constraint_common}
1, &~~{{\sum}_{k=1}^K} c_k \leq \mathop {\min} \limits_{k} ~ R^{\rm (C)}_k \text{and} ~\forall k \in \mathcal{K},\\
0, &~~{{\sum}_{k=1}^K} c_k > \mathop {\min} \limits_{k} ~ R^{\rm (C)}_k \text{and} ~\forall k \in \mathcal{K},
\end{aligned}\right.\\
\Gamma_{\rm {QoS}}&=\left\{
\begin{aligned}\label{eq_constraint_QoS}
1, &~~c_k+R_k^{\rm (P)} \geq \zeta_k ~ \text{and} ~ \forall k \in \mathcal{K},\\
0, &~~c_k+R_k^{\rm (P)} < \zeta_k ~ \text{and} ~ \forall k \in \mathcal{K},
\end{aligned}\right.\\
\Gamma_{\rm {linear}}&=\left\{
\begin{aligned}\label{eq_constraint_Linear}
1, &~~{{\sum}_{k=0}^K}{|v_{k,l}|}\leq\Delta ~ \text{and} ~ \forall l \in \mathcal{L},\\
0, &~~{{\sum}_{k=0}^K}{|v_{k,l}|}>\Delta ~ \text{and} ~ \forall l \in \mathcal{L}.
\end{aligned}\right.
\end{align}

\begin{remark}
The penalty method is widely employed to ensure a feasible solution, but it may lead to sparse rewards and high variance\cite{10480579}. To balance learning efficiency and {to meet the} constraints, we introduce penalty terms for no-convex constraints, i.e., (\ref{P0_C1}), (\ref{P0_C2}), (\ref{P0_C3}), and (\ref{P0_C4}), while simple constraints, i.e., (\ref{P0_C5}), (\ref{P0_C6}), (\ref{P0_C7}), (\ref{P0_C8}), and (\ref{P0_C9}), are {met} through activation functions.
\end{remark}

\subsection{DS-PPO Approach Framework}

\begin{figure*}[!htb]
\centering
\includegraphics[width=0.75\linewidth]{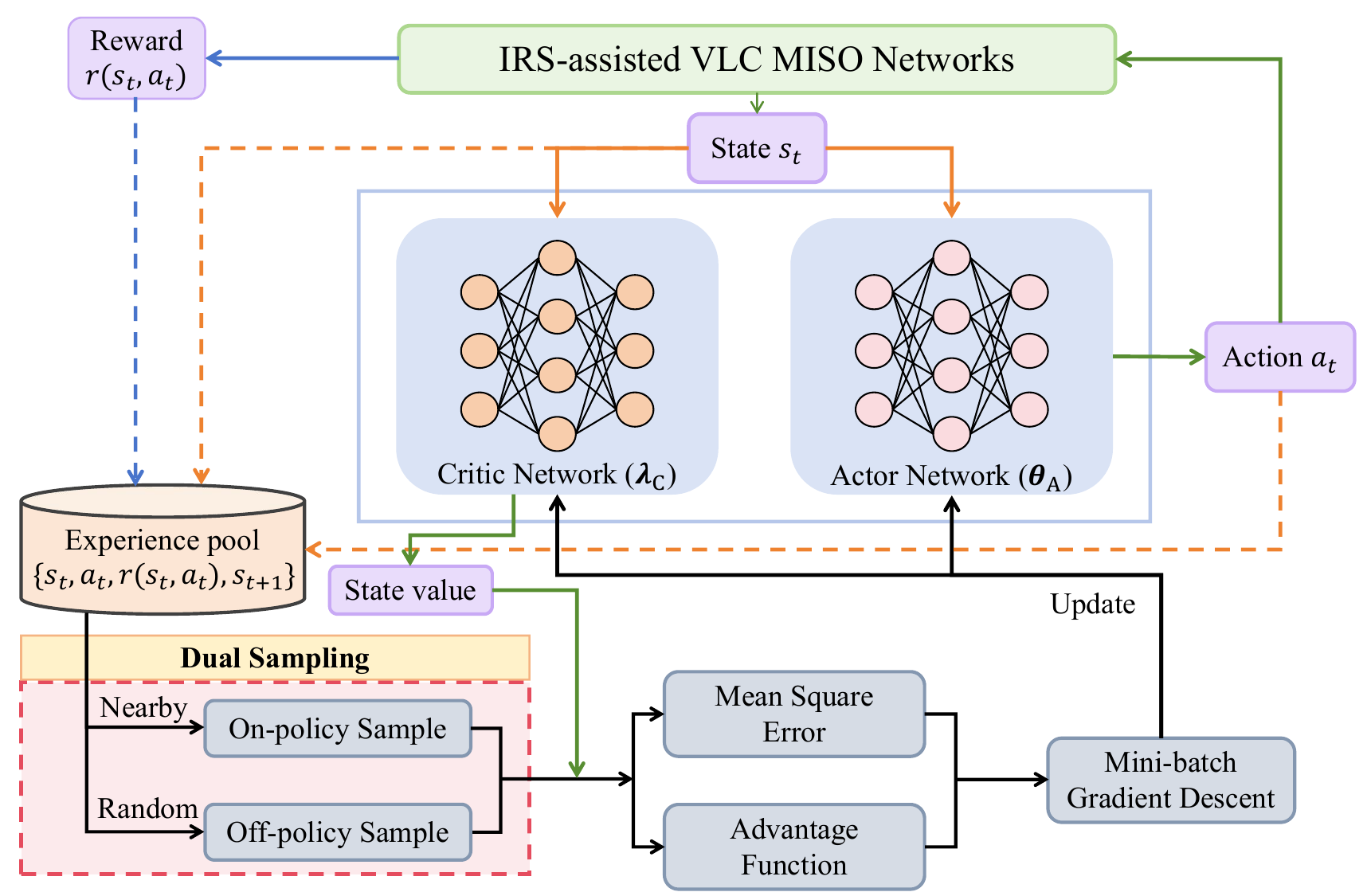}
\caption{{Framework of the proposed DS-PPO Approach. The IRS-assisted VLC MISO network utilizes dual sampling to generate on-policy and off-policy samples, which are processed through actor-critic networks to optimize actions via gradient descent.}}
\label{markov_decision_process}
\end{figure*}

In practice, PPO is a model-free, on-policy method that integrates policy gradient (PG) with an actor-critic (AC) framework\cite{10032267}, and is well-suited for obtaining an effective solution to NP-hard optimization problems as in $\boldsymbol{\mathcal P_1}$. The key advantage of PPO is its monotonic policy improvement via a clipped surrogate objective function, which limits the update step of the parameters to avoid performance collapse\cite{schulman2017proximal}. As a result, the robustness and stability of PPO would be guaranteed. However, due to the low sample diversity caused by on-policy sampling and the poor ability of return-to-go (RTG) in controlling the bias and variance, the exploration efficiency of PPO is always diminished in handling complicated problems.

To address this issue, we propose DS-PPO, an enhanced version of PPO with two key improvements. First, DS-PPO incorporates off-policy sampling alongside on-policy sampling, which greatly enhances the diversity of samples. Second, the generalized advantage estimation (GAE) method is adopted in DS-PPO to {improve} the policy gradient estimation in controlling the bias and variance. Fig.~\ref{markov_decision_process} illustrates the framework of the proposed DS-PPO approach. In DS-PPO, $\boldsymbol\theta_{\rm A}$ and $\boldsymbol\lambda_{\rm C}$ are the parameters of the actor network and critic network, respectively. Denote $\pi_{\boldsymbol\theta_{\rm A}}$ and $V_{\boldsymbol\lambda_{\rm C}}(s_t)$ as the policy of action selection and the state-value estimation function, respectively. The surrogate objective function of DS-PPO is defined as
\begin{equation}\label{eq_ob_func}
\mathcal{L}(\boldsymbol \theta_{\rm A})=\mathbb{E}\left[\frac{\pi_{\boldsymbol\theta_{\rm A}}(a_t|s_t)}{\pi_{\boldsymbol\theta_{\rm A}^{\rm old}}(a_t|s_t)} \hat{A}_t\right]=\mathbb{E}\left[\rho_t(\boldsymbol \theta_{\rm A})\hat{A}_t\right],
\end{equation}
where $\hat{A}_t$ is the advantage estimation function, and $\rho_t(\boldsymbol \theta_{\rm A})={\pi_{\boldsymbol\theta_{\rm A}}(a_t|s_t)}/{\pi_{{\boldsymbol\theta_{\rm A}^{\rm {old}}}}(a_t|s_t)}$ is the probability ratio. Particularly, $\rho_t(\boldsymbol \theta_{\rm A}) > 1$ indicates that the current action is better than the previous one. Conversely, $0<\rho_t(\boldsymbol \theta_{\rm A})<1$ means the current action is less reliable.

In face, the traditional advantage estimation function is given by
\begin{equation}
\tilde{A}^{\text{traditional}}_t = r_t + V_{\boldsymbol\lambda_{\rm C}}(s_{t+1}) - V_{\boldsymbol\lambda_{\rm C}}(s_t),
\end{equation}
where $r_t$ is the reward obtained at time step $t$, $V_{\boldsymbol\lambda_{\rm C}}(s_{t+1})$ and $V_{\boldsymbol\lambda_{\rm C}}(s_t)$ are the state values estimated by the critic network at time step $t+1$ and $t $, respectively.

Unlike the traditional advantage estimation function, GAE combines Monte Carlo and temporal difference error to balance the bias and the variance of policy gradient estimation, providing stable performance during policy updates\cite{schulman2015high}. Specifically, in GAE, the temporal difference error at the $t$-th time step is first calculated by
\begin{equation}
\delta_t = r_t + \gamma V_{\boldsymbol\lambda_{\rm C}}(s_{t+1}) - V_{\boldsymbol\lambda_{\rm C}}(s_t),
\end{equation}
where $\gamma$ is the discount factor. Then, starting from the last time step $T-1$, GAE at the $t$-th time step can be recursively calculated by
\begin{equation}\label{eq_gae}
\hat{A}_t = \delta_t + \gamma \hat{A}_{t+1},~t=T-2,\ldots,1,0,
\end{equation}
where $\hat{A}_{T-1}$ is set as $\delta_{T-1}$.

To ensure that the updated policy $\pi_\theta(a_t|s_t)$ satisfies the trust region constraint and avoids excessively large update step, the surrogate objective function in (\ref{eq_ob_func}) can be converted into
\begin{equation}\label{eq_clip_ob_func}
\begin{aligned}
&\mathcal{L}_{\rm {CLIP}}(\boldsymbol\theta_{\rm A})\\
&=\mathbb{E}\left[\min(\rho_t(\boldsymbol\theta_{\rm A})\hat{A}_t,\text{clip}(\rho_t(\boldsymbol\theta_{\rm A}),1-\epsilon,1+\epsilon)\hat{A}_t)\right],
\end{aligned}
\end{equation}
where clip($\cdot$) is exploited to restrict the range of the update step, preventing oscillations caused by excessive updates.

\begin{algorithm}[htb!]
\footnotesize
\SetKwInOut{Input}{Input}\SetKwInOut{Output}{Output}\SetKwInOut{Initialize}{Initialize}
\caption{The Proposed DS-PPO Approach}
\label{alg:alg DS-PPO}
\Input{Channel gains $\{\boldsymbol{h}_k^{\rm {(1)}},\boldsymbol{h}_k^{\rm {(2)}},\boldsymbol{h}_{\rm E}^{\rm {(1)}}\}$, on-policy batch size $b_{\rm {on}}$, off-policy batch size $b_{\rm {off}}$, max step $\mathcal T$;}
\Output{Action $\boldsymbol a$, State $\boldsymbol s$, $P_{\rm T}$, $\{R_k^{\rm (C)}\}$, $\{R_k^{\rm (P)}\}$, $R_{\rm E}^{\rm (C)}$, $\{R_{\rm E}^{\rm (P)}\}$, and SEE;}
\Initialize{Policy parameters $\boldsymbol \theta_A$, value parameters $\boldsymbol \lambda_C$, experience pool $\boldsymbol{\mathcal{B}}$, learning rates $\alpha_{\rm A}$ and $\alpha_{\rm C}$;}
Obtain the initial state $s_0$ of the environment\;
\For{{\rm step} $t=1$ {\rm to} $\mathcal T$}{
Take Action $a_t$ by the policy $\pi_{\boldsymbol\theta_{\rm {A}}}(a_t|s_t)$ based on the current state $s_t$\;
Compute the immediate reward $r_t$ by (\ref{eq_reward})\;
Observe the next state $s_{t+1}$\;
Store $\{{a}_t, {s}_t, {r}_t, {s}_{t+1}\}$ into the experience pool $\boldsymbol{\mathcal{B}}$\;
\tcp{\textrm{\textbf{on-policy update}}}
Sample $b_{\rm {on}}$ transitions from experience pool $\boldsymbol{\mathcal{B}}$\;
\For{{\rm epoch} $e_{\rm on}=1$ {\rm to} $\mathcal{N}_{\rm on}$}{
Compute $\mathcal{L}_{\rm {CLIP}}(\boldsymbol \theta_{\rm A})$ according to (\ref{eq_clip_ob_func})\;
Update $\boldsymbol \theta_{\rm A}^{\rm {old}}\leftarrow\boldsymbol \theta_{\rm A}$ following (\ref{eq_update_A})\;
Compute $\mathcal{L}_{\rm {MSE}}(\boldsymbol \lambda_{\rm C})$ according to (\ref{eq_loss})\;
Update $\boldsymbol \lambda_{{\rm C}}^{\rm {old}}\leftarrow\boldsymbol \lambda_{\rm C}$ following (\ref{eq_update_C})\;
}
\tcp{\textrm{\textbf{off-policy update}}}
Sample $b_{\rm {off}}$ transitions from experience pool $\boldsymbol{\mathcal{B}}$\;
\For{{\rm epoch} $e_{\rm off}=1$ {\rm to} $\mathcal{N}_{\rm off}$}{
Compute $\mathcal{L}_{\rm {CLIP}}(\boldsymbol \theta_{\rm A})$ according to (\ref{eq_clip_ob_func})\;
Update $\boldsymbol \theta_{\rm A}^{\rm {old}}\leftarrow\boldsymbol \theta_{\rm A}$ following (\ref{eq_update_A})\;
Compute $\mathcal{L}_{\rm {MSE}}(\boldsymbol \lambda_{\rm C})$ according to (\ref{eq_loss})\;
Update $\boldsymbol \lambda_{{\rm C}}^{\rm {old}}\leftarrow\boldsymbol \lambda_{\rm C}$ following (\ref{eq_update_C})\;
}
Update state $s_t \leftarrow s_{t+1}$\;
}
\end{algorithm}

DS-PPO achieves monotonic policy improvement by updating the parameters of actor network $\boldsymbol\theta_{\rm A}$, based on the gradient of $\mathcal{L}_{\rm {CLIP}}(\boldsymbol\theta_{\rm A})$. Specifically, $\boldsymbol\theta_{\rm A}$ is optimized using mini-batch stochastic gradient descent (SGD), where transitions $\{\boldsymbol{a}_t, \boldsymbol{s}_t, \boldsymbol{r}_t, \boldsymbol{s}_{t+1}\}$ are sampled from experience pool $\boldsymbol{\mathcal{B}}$. That is, $\boldsymbol\theta_{\rm A}$ can be given by
\begin{equation}\label{eq_update_A}
\begin{aligned}
\boldsymbol\theta_{\rm A}={\boldsymbol \theta }^\mathrm{old}_\mathrm{A} - \alpha _\mathrm{A}\left[\frac{1}{B}{\sum}_{j=1}^{B}\left({ \nabla _{{\boldsymbol \theta }_\mathrm{A}}{\mathcal{L}}_{\rm {CLIP}}(\boldsymbol\theta_{\rm A})}\right)\right],
\end{aligned}
\end{equation}
where $\alpha_{\rm A}$ is the learning rate of actor network, and $B \in \{b_{\rm on},b_{\rm off}\}$ is the number of transitions sampled from experience pool, with $b_{\rm on}$ and $b_{\rm off}$ denoting the number of transitions sampled by on-policy strategy and off-policy strategy, respectively. For updating $\boldsymbol\lambda_{\rm C}$, the mean squared error (MSE) is exploited as the loss function, which is given by
\begin{equation}\label{eq_loss}
\mathcal{L}_{\rm {MSE}}(\boldsymbol\lambda_{\rm C})=\left[V_{\boldsymbol\lambda_{\rm C}}(\boldsymbol s_t)-V_{\rm {target}}(\boldsymbol s_t)\right]^2,
\end{equation}
where $V_{\rm {target}}(\boldsymbol s_t)$ represents the target state-value function. Additionally, the mini-batch SGD method is also employed to update $\boldsymbol\lambda_{\rm C}$ as follows:
\begin{equation}\label{eq_update_C}
{\boldsymbol \lambda }_\mathrm{C} = {\boldsymbol \lambda }^\mathrm{old}_\mathrm{C} - \alpha _\mathrm{C}\left[\frac{1}{B}{\sum}_{j=1}^{B} \nabla _{{\boldsymbol \lambda }_\mathrm{C}}\mathcal{L}_{\rm {MSE}}(\boldsymbol\lambda_{\rm C})\right],
\end{equation}
where $\alpha_{\rm C}$ denotes the learning rate of critic network.

With the DS-PPO approach, in the $t$-th time step, the agent first takes action $a_t$ based on the current state $s_t$ according to the policy $\pi_{\boldsymbol\theta_{\rm A}}$. Then, after the interaction between the agent and the environment, corresponding immediate reward $r_t$ and the next state $s_{t+1}$ are obtained. Meanwhile, the transition $\{a_t, s_t, r_t, s_{t+1}\}$ obtained by the interaction is stored into the experience pool. Next, $b_{\rm on}$ and $b_{\rm off}$ transitions are sampled from the experience pool by on-policy strategy and off-policy strategy, respectively. After that, the value of the advantage estimation function is calculated with (\ref{eq_gae}). Finally, the parameters of the neural networks, i.e., $\boldsymbol\theta_{\rm A}$ and $\boldsymbol\lambda_{\rm C}$, are respectively updated with (\ref{eq_update_A}) and (\ref{eq_update_C}). In summary, Algorithm~\ref{alg:alg DS-PPO} presents the detailed process of the proposed DS-PPO approach.

\begin{remark}
The computational complexity of the proposed DS-PPO approach, which depends on the scale of neural networks\cite{10107766}, is about $\mathcal{O}({\sum}_{p=1}^{P} m_{p-1}\cdot m_p)$, where $P$ and $m_p$ are the number of the layers in neural networks and the number of neural units in the $p$-th layer, respectively.
\end{remark}

\section{Simulation Results}
\label{sec:5_simulation_results}
\begin{table}[h!]
\centering
\footnotesize
\caption{Simulation Settings}
\label{tbl:simulation_settings}
\begin{tabular}{l|c|c}
  \hline\hline
  \textbf{Description} & \textbf{Symbol} & \textbf{Value} \\ \hline\hline
  Optical filter gain & $g_{\rm of}$ & 1 \\ 
  Refractive index & $\kappa$ & 1.5\cite{10433073} \\ 
  Field of view & $\Psi_{\rm F}$ & $75^\circ$\cite{10552701} \\ 
  Half intensity view angle & $\Omega_{1/2}$ & $60^\circ$ \\ 
  Physical area of PD & $A_{\rm P}$ & 1 cm$^2$\cite{10433073} \\
  Maximum drive current & $I_{\rm max}$ & 5 A\cite{10552701,1004299} \\
  Minimum drive current & $I_{\rm min}$ & 0 A\cite{10552701,1004299} \\ 
  Forward voltage of LEDs & $U_{\rm LED}$ & 2 V \\ 
  Constant circuit power consumption & $P_{\rm C}$ & 2 W \\
  Power budget & $P_{\rm max}$ & 20 W \\
  QoS threshold & $\zeta$ & 2 bit/s/Hz \\
  Noise power density & $\sigma_k^2,\sigma_{\rm E}^2$ & $10^{-13}$ W\cite{10552701} \\ 
  Reflection coefficient of IRS elements & $\rho$ & 0.9\cite{10024150} \\ 
  \hline\hline
\end{tabular}
\end{table}
In this section, simulation results and analyses of the proposed DS-PPO approach are presented. In the simulations, the considered VLC network is deployed in a room with size of 6 m $\times$ 6 m $\times$ 3 m, where 6 LEDs, an IRS with 16 reflecting elements, 2 LUs, and an Eve are involved. LEDs are evenly distributed on the ceiling, whose coordinations are (1.5, 2, 3) m, (3, 2, 3) m, (4.5, 2, 3) m, (1.5, 4, 3) m, (3, 4, 3) m and (4.5, 4, 3) m, respectively. Moreover, the IRS is mounted on one of the walls, while the LUs and the Eve are randomly located on the ground. The remaining simulation settings are summarized in Table \ref{tbl:simulation_settings}.

\begin{table}[h!]
\centering
\footnotesize
\caption{DS-PPO Hyper-parameters Settings}
\label{tbl:ds-ppo_hyperparameters}
\begin{tabular}{l|c}
  \hline\hline
  \textbf{System Parameter} & \textbf{Value} \\ \hline\hline
  Learning rate of actor network $\alpha_{\rm A}$ & $2.5\times10^{-4}$ \\ 
  Learning rate of critic network $\alpha_{\rm C}$ & $2.5\times10^{-4}$ \\ 
  Optimizer of neural networks & Adam \\ 
  On-policy training epoch $\mathcal N_{\rm on}$ & 10 \\
  Off-policy training epoch $\mathcal N_{\rm off}$ & 3 \\
  Training steps $\mathcal T$ & $1.5\times10^7$\cite{10032267} \\
  Buffer size & 40960\cite{10032267} \\ 
  On-policy batch size $b_{\rm {on}}$ & 2048 \\ 
  Off-policy batch size $b_{\rm {off}}$ & 256 \\ 
  Discount factor $\gamma$ & 0\cite{10032267} \\
  Clip factor $\epsilon$ & 0.2\cite{schulman2017proximal} \\
  \hline\hline
\end{tabular}
\end{table}
The proposed DS-PPO approach utilizes two neural networks, i.e., the actor (policy network) and the critic (value network), both of which are structured as four-layer feed-forward deep neural networks, i.e., an input layer, two hidden layers, and an output layer. Each hidden layer comprises 256 neurons and the hyperbolic tangent function is adopted as the activation function. For the actor network, the input dimension is  $NLK+5K+L+3$, and the output dimension is $NLK+2K+L+1$. Similarly, the critic network shares the same input dimension as the actor network (i.e., $NLK+5K+L+3$), while its output dimension is 1. The remaining hyper-parameters setting are summarized in Table \ref{tbl:ds-ppo_hyperparameters}. Moreover, the learning rates listed in the table are the initial values, which follows an exponential decay, i.e., $\alpha(t)=\alpha_0 \times 10^{-{t}/{T_{\rm decay}}}$, where $\alpha_0$, $t$, and $T_{\rm decay}$ are the initial learning rate, the time step, and the decay step, respectively.
\subsection{Convergence}
\begin{figure}[htb!]
  \centering
  \includegraphics[width=0.475\textwidth]{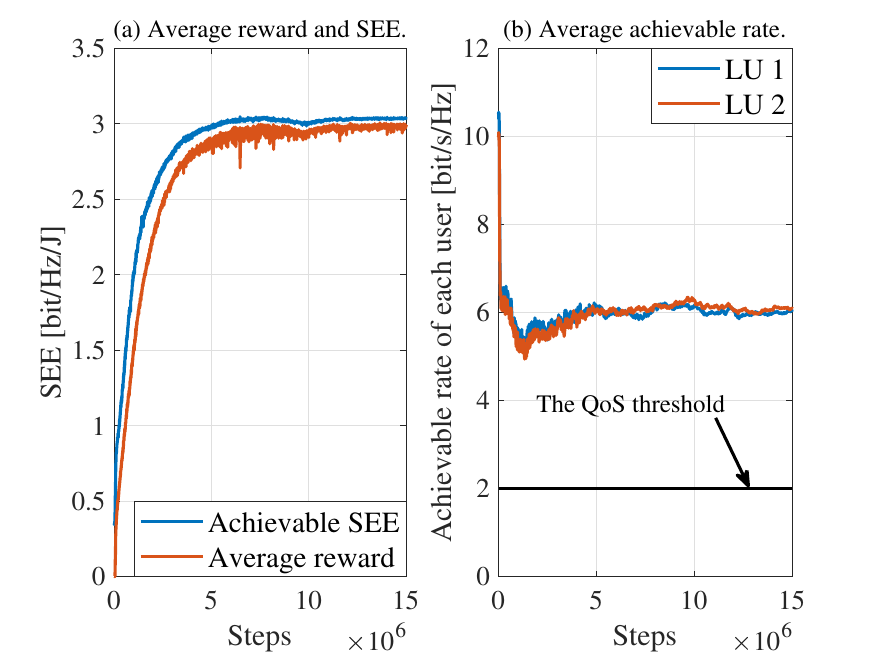}
  \caption{Convergence behaviors of the DS-PPO approach.}
  \label{fig:convergence_ds-ppo}
\end{figure}

Fig.~\ref{fig:convergence_ds-ppo} illustrates the training process curves of the proposed DS-PPO approach to show its convergence behaviors. To clearly present the overall training trend, moving average method is adopted to smooth the curves. In Fig.~\ref{fig:convergence_ds-ppo}\textcolor{blue}{a}, the curves of the achievable SEE and average reward versus the training steps are plotted. It is seen that both the achievable SEE and average reward converge as the number of training steps increases, and the gap between them decreases over time. This is because the quality of samples in the experience pool of the proposed DS-PPO approach gradually optimizes as the number of training steps increasing. Besides, there are slight jitters in the reward curve which are due to the exploration mechanism by adding random exploration noise during each action selection process\cite{10032267}. Fig.~\ref{fig:convergence_ds-ppo}\textcolor{blue}{b} plots the training curves of the achievable rate of each LU. It is shown that the achievable rate of each LU is always above the required QoS threshold. 
\subsection{Selection of Learning Rate}
\begin{figure}[htb!]
  \centering
  \includegraphics[width=0.475\textwidth]{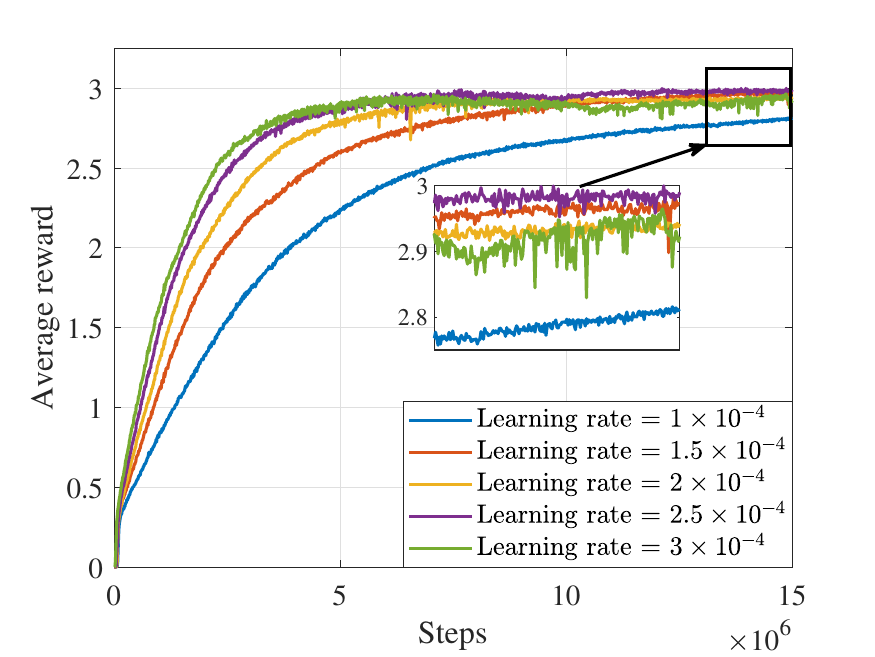}
  \caption{Convergence performances of different learning rates.}
  \label{fig:lr}
\end{figure}
 In DRL, the learning rate determines the step size for each parameter update and significantly impacts the training speed. Particularly, a larger learning rate can cause significant fluctuations, leading the average reward unable to converge; In contrast, a smaller learning rate may result in excessively slow convergence, potentially leading to underfitting. Thus, it is crucial to evaluate the performance of the DS-PPO algorithm under different learning rates. 
As shown in Fig.~\ref{fig:lr}, when the learning rate is set to $1\times10^{-4}$, the average reward grows slower with the increment of the number of steps compared to those with larger learning rate. Moreover, at a learning rate of $3\times10^{-4}$, the average reward fails to ensure a better convergence. This is because the neural networks overlearn the exploration noise during each action selection process, leading to overfitting. The average reward for the remaining curves are relatively better, with optimal performance observed at a learning rate of $2.5\times10^{-4}$, and thus selected for subsequent simulations.
\subsection{Comparison of the DS-PPO Approach with Baselines}
\label{sec:sec:baseline_methods}
\begin{figure}[htb!]
  \centering
  \includegraphics[width=0.475\textwidth]{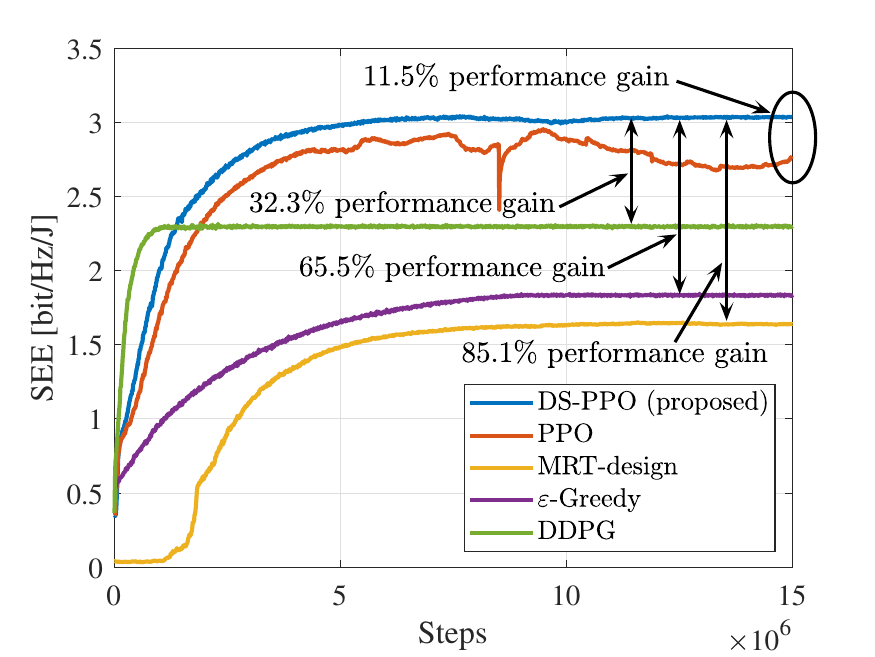}
  \caption{SEE achieved with different approaches.}
  \label{fig:comparison_reward}
\end{figure}

\begin{figure}[htb!]
  \centering
  \includegraphics[width=0.475\textwidth]{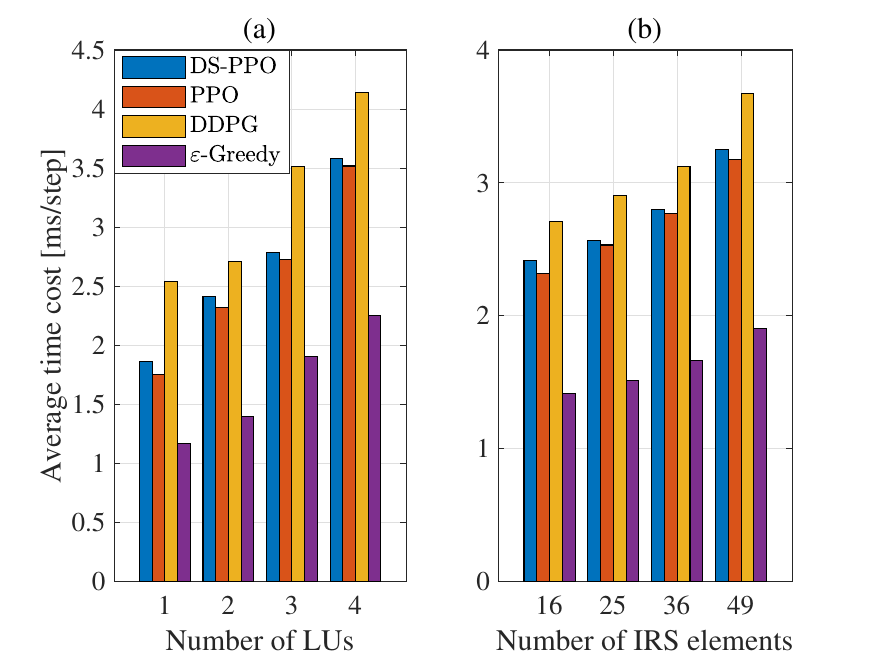}
  \caption{Average time cost with different baseline approaches.}
  \label{fig:comparison_time}
\end{figure}

To show the advantages of the proposed DS-PPO algorithm in addressing complex optimization problems, Fig.~\ref{fig:comparison_reward} and Fig.~\ref{fig:comparison_time} compare the average reward and the average time cost of the proposed DS-PPO approach with four widely adopted baseline approaches, i.e.,
\begin{itemize}
  \item[$\bullet$] \textbf{The PPO-based approach:} The PPO-based approach with on-policy sampling presented in \cite{10679152} is adopted as a baseline approach.
  \item[$\bullet$] \textbf{The DDPG-based approach:} The DDPG-based approach with discount rate $\gamma=0.9$ and soft update coefficient $\tau=0.05$ presented in \cite{9667214} is adopted as a baseline.
  \item[$\bullet$] \textbf{The $\varepsilon$-Greedy-based approach:} In the $\varepsilon$-Greedy-based approach, the agent explores the action to maximize the achievable SEE in every time step.
  \item[$\bullet$] \textbf{The MRT-design:} Referring to \cite{8968350}, only the MRT approach is adopted to design an efficient beamforming direction.
\end{itemize}

\subsubsection{Performance comparison}
Fig.~\ref{fig:comparison_reward} illustrates the achieved SEE with different approaches. It is observed that the proposed DS-PPO approach achieves a higher SEE upon final convergence compared to these four baseline approaches, indicating that a significant performance improvement can be achieved by our proposed DS-PPO approach. Particularly, the SEE achieved by DS-PPO is about $11.5\%$ higher than that achieved by PPO. This is because DS-PPO adopts the advanced GAE methods, which can effectively degrade the variance and controls the bias of policy gradient estimation. Moreover, the proposed DS-PPO converges faster than PPO, because it combines the off-policy sampling strategy with on-policy sampling strategy to update the parameters of neural networks. Besides, it is {important to note} that although DDPG achieves remarkable advantages in terms of the faster convergence speed, the corresponding SEE is relatively low compared with DS-PPO and PPO approaches.

\subsubsection{Time-cost comparison}
Time cost is also an important metric in evaluating the performance of algorithms. To compare the time cost of the proposed DS-PPO approach with the baseline approaches under different scenarios, Fig.~\ref{fig:comparison_time}\textcolor{blue}{a} and Fig.~\ref{fig:comparison_time}\textcolor{blue}{b} plot the average time cost with different numbers of LUs and IRS elements, respectively. It can be seen that $\varepsilon$-Greedy exhibits the lowest time cost as it does not involve neural networks. Moreover, DDPG exhibits the longest time cost due to its reliance on a pair of AC networks and a pair of target AC networks, resulting in high computational complexity. Notably, because of the off-policy sampling strategies, DS-PPO sacrifices less than $5\%$ of the time cost compared to PPO while achieving $12.1\%$ performance gain in terms of the average reward demonstrated in Fig.~\ref{fig:comparison_reward}.
\begin{figure}[h!]
  \centering
  \includegraphics[width=0.475\textwidth]{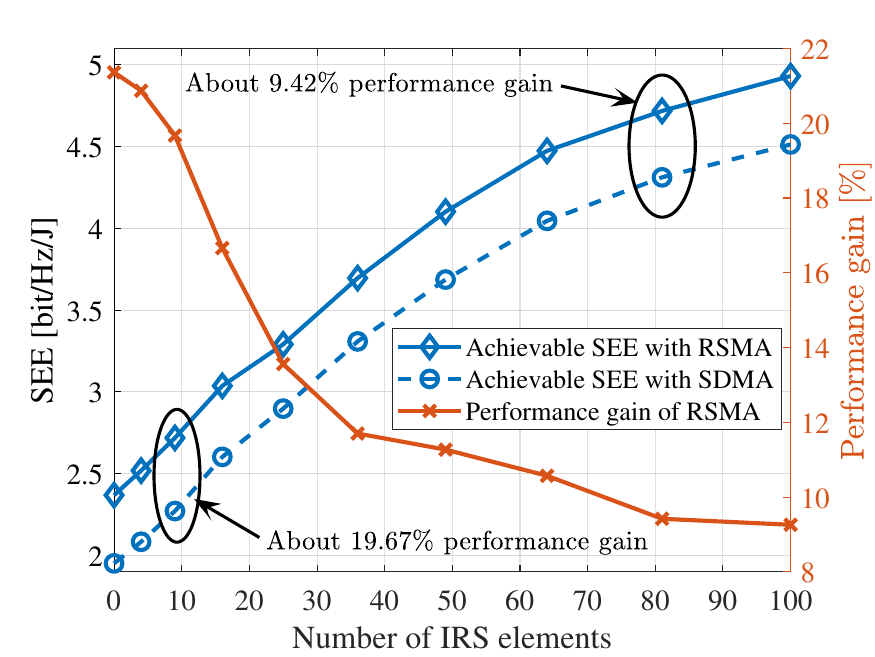}
  \caption{Comparison of SEE achieved by RSMA with SDMA.}
  \label{fig:SEE_N}
\end{figure}
\subsection{Effects of the Number of IRS Elements}
Fig.~\ref{fig:SEE_N} compares the SEE achieved by RSMA with that achieved by SDMA versus the number of IRS elements. It can be seen that with the increment of the number of IRS elements, the {SEEs} achieved by RSMA and SDMA grow gradually. The reason is that more IRS elements can provide more spatial DoF to mitigate the IUI better. 
Moreover, the SEE achieved by RSMA scheme consistently outperforms that achieved by SDMA, although the performance gap diminishes with the growing of the number of IRS elements. This is due to the SIC and the flexible rate allocation of RSMA, which enhances spectral efficiency and interference management.

\subsection{Effects of the Number of LUs}
\begin{figure}[htb!]
  \centering
  \includegraphics[width=0.475\textwidth]{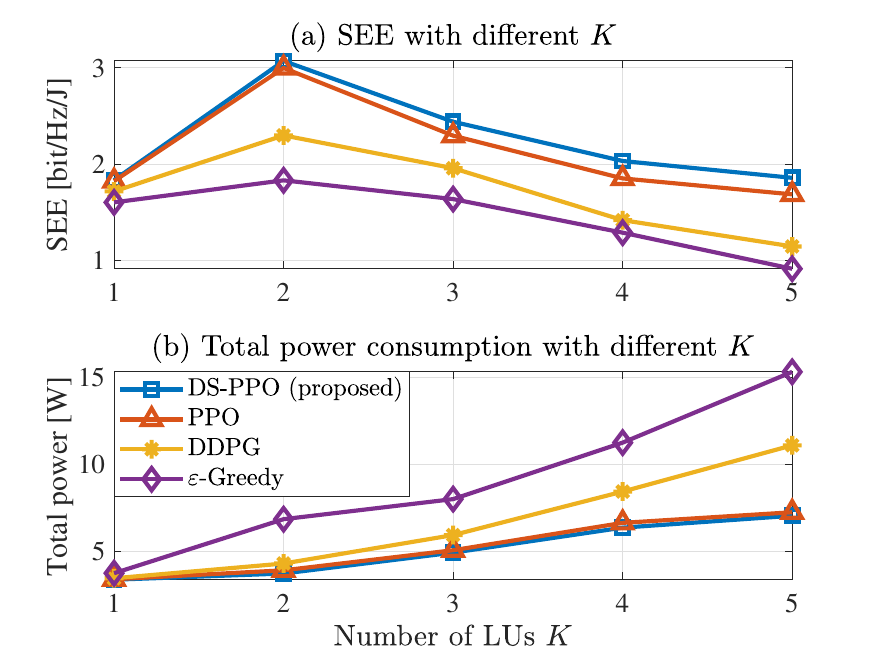}
  \caption{SEE and total power with different number of LUs $K$.}
  \label{fig:comparison_K}
\end{figure}

{The number of LUs is varied to analyze the system performance.} Fig.~\ref{fig:comparison_K} simulates the achievable SEE and the total power consumption of the network versus the number of LUs. It can be observed from Fig.~\ref{fig:comparison_K}\textcolor{blue}{a} that as the number of LUs grows, the achievable SEE first increases and then decreases. This is because when the number of LUs is relatively small, as the number of LUs grows, SEE increases due to the advantage of the DoF provided by multiple LEDs. However, when the number of LUs is larger than a specific value, more LUs will lead to more severe IUI, which in turn reduces the SEE. Moreover, since the increasing IUI leads the system to consume more power to satisfy the more stringent QoS requirements, it can be seen from Fig.~\ref{fig:comparison_K}\textcolor{blue}{b} that the total consumed power grows with the increasing of the number of LUs.
\subsection{Effects of Total Power Budget}
\begin{figure}[htb!]
  \centering
  \includegraphics[width=0.475\textwidth]{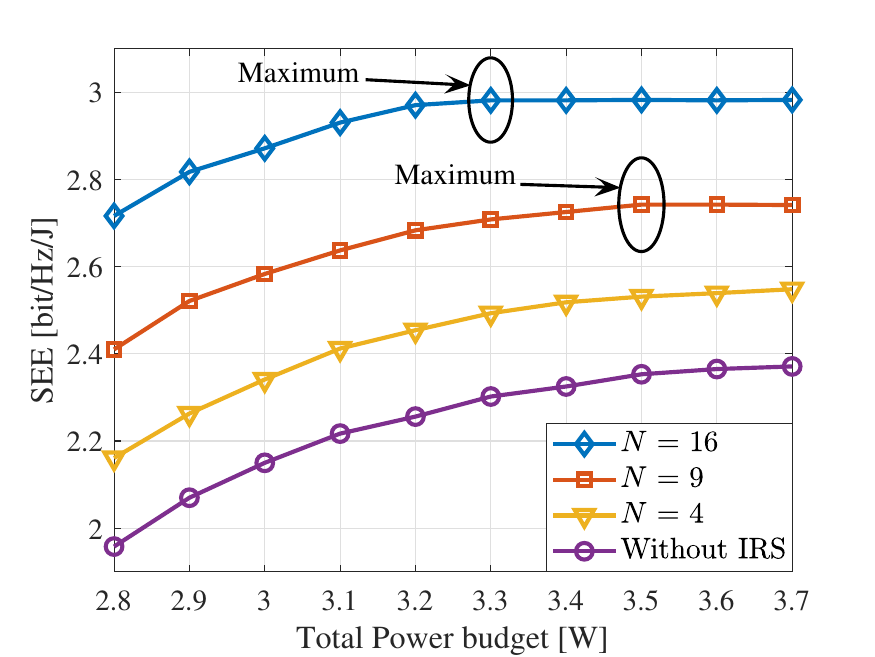}
  \caption{SEE versus total power budget.}
  \label{fig:SEE_power}
\end{figure}
Fig.~\ref{fig:SEE_power} illustrates the effects of the total power budget on SEE with different numbers of IRS elements. It can be seen that with the increment of the power budget, SEE first grows and then stabilizes. This is because when the system operates in its power-limited region, the achievable SEE increases gradually with the growth of the transmit power budget. However, when the power budget {approaches} a relatively large value, SEE would no longer keep increasing due to the linear operating region limitation of LEDs. Moreover, Fig.~\ref{fig:SEE_power} reveals that more IRS elements lead to a higher achievable SEE, which is consistent with the phenomena illustrated in Fig.~\ref{fig:SEE_N}.
\subsection{Effects of QoS Threshold}
\begin{figure}[htb!]
  \centering
  \includegraphics[width=0.475\textwidth]{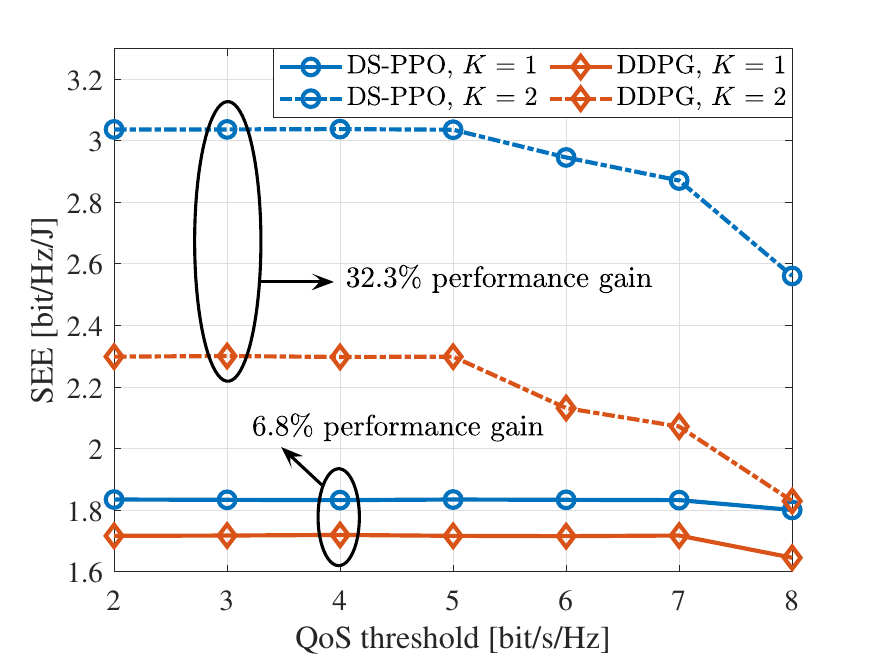}
  \caption{SEE versus the QoS threshold.}
  \label{fig:SEE_QoS}
\end{figure}
Fig.~\ref{fig:SEE_QoS} shows SEE achieved by DS-PPO and DDPG versus the QoS threshold with the number of LUs $K = \{1, 2\}$, respectively. It can be observed that as the QoS threshold increases, the achievable SEE remains constant at the beginning and then gradually decreases. The reason is that when the QoS threshold is relatively small, e.g., less than 7 bit/s/Hz for $K=1$, the total consumed power is fixed, leading the SEE to remain unchanged. Meanwhile, when the QoS threshold grows beyond a certain value, the total consumed power sharply increases, and reduces the SEE. Moreover, it can be seen that the achievable SEE when $K = 2$ is higher than that when $K = 1$, which is consistent with that in Fig.~\ref{fig:comparison_K}. Besides, Fig.~\ref{fig:SEE_QoS} shows that DS-PPO outperforms DDPG by $32.1\%$ when $K=2$, which reveals that our proposed DS-PPO approach has significant advantages of achieving great gain with respect to the traditional DDPG approach.
\section{Conclusion}
\label{sec:6_conclusion}
This paper {has} studied the IRS-assisted VLC MISO network with RSMA. To maximize the SEE of the system, we {have} formulated a problem to jointly optimize beamforming vectors, common information rate, direct current bias, and alignment matrices between the IRS elements and LEDs as well as the IRS elements and LUs, while satisfying the constraints of the total power budget, the common information rate allocation, the QoS threshold, and the linear operating region of LEDs. Then, a DRL-based DS-PPO approach with dual sample strategies, i.e., on-policy and off-policy samples, {has been} proposed to handle the problem by adopting the advanced GAE method. Simulation results {have} demonstrated the advantages of the proposed DS-PPO approach to traditional baseline approaches like PPO, and there exists a maximum SEE {with respect to} the number of LUs. Moreover, compared to the traditional multiple access schemes and networks without IRS deployment, the RSMA scheme and the deployment of IRS significantly have improved the SEE performance. 

\bibliographystyle{IEEEtran}
\bibliography{IEEEexample_abrv}

\end{CJK}
\end{document}